\documentclass[12pt]{article}
\pdfoutput=1
\usepackage{jheppub}
\usepackage{bm}
\usepackage{amssymb}
\title{Inertial frames without the relativity principle}
\author{Valentina Baccetti,  Kyle Tate, \textmd{and} Matt Visser}
\affiliation{School of Mathematics, Statistics, and Operations Research, \\
Victoria University of Wellington, PO Box 600, Wellington 6140, New Zealand}
\emailAdd{valentina.baccetti@msor.vuw.ac.nz}
\emailAdd{kyle.tate@msor.vuw.ac.nz}
\emailAdd{matt.visser@msor.vuw.ac.nz}
\abstract{
Ever since the work of von Ignatowsky circa 1910 it has been known (if not always widely appreciated) that the relativity principle, combined with the basic and fundamental physical assumptions of locality,  linearity,  and isotropy, leads almost uniquely to either the Lorentz transformations of special relativity or to Galileo's transformations of classical Newtonian mechanics.  Consequently, if one wishes (for whatever reason) to entertain the possibility of Lorentz symmetry breaking within the context of the class of local physical theories, then it seems likely that one will have to abandon (or at the very least grossly modify) the relativity principle.   

Working within the framework of local physics, we reassess the notion of spacetime transformations between inertial frames in the absence of the relativity principle, arguing that significant and nontrivial physics can still be extracted as long as the transformations are at least linear. An interesting technical aspect of the analysis is that the transformations now form a \emph{groupoid}/\emph{pseudo-group}  --- it is this technical point that permits one to evade the von Ignatowsky argument.  

Even in the absence of a relativity principle we can (assuming locality and linearity) nevertheless deduce clear and compelling rules for the transformation of space and time, rules for the composition of 3-velocities, and rules for the transformation of energy and momentum. Within this framework, the energy-momentum transformations are in general \emph{affine}, but may be \emph{chosen} to be linear, with the 4-component vector $P = (E,-\bm{p}^T)$ transforming as a row vector, while the 4-component vector of space-time position $X=(t,\bm{x}^T)^T$ transforms as a column vector. 

As part of the analysis we identify two particularly elegant and physically compelling models implementing  ``minimalist'' violations of Lorentz invariance --- in the first of these minimalist models all Lorentz violations are confined to carefully delineated particle physics sub-sectors, while the second minimalist Lorentz-violating model depends on one free function of absolute velocity, but otherwise preserves as much as possible of standard Lorentz invariant physics. 

\noindent
7 December 2011; 12 December 2011;  28 March 2012; \LaTeX-ed \today
}
\begin{document}
\maketitle
\def\x{\bm{x}}
\def\p{\bm{p}}
\def\0{\bm{0}}
\def\n{\bm{n}}
\def\vv{\bm{v}}
\def\u{\bm{u}}
\def\w{\bm{w}}
\def\bpi{\bm{\pi}}
\def\blambda{\bm{\lambda}}
\def\P{{\mathcal{P}}}
\def\C{{\mathcal{C}}}
\def\E{{\mathcal{E}}}
\def\R{{\mathbb{R}}}
\section{Introduction}
\label{S:intro}
\def\x{\bm{x}}
\def\p{\bm{p}}
\def\0{\bm{0}}
\def\n{\bm{n}}
\def\v{\bm{v}}
\def\u{\bm{u}}
\def\w{\bm{w}}
\def\bpi{\bm{\pi}}

In 1910 von Ignatowsky established a very tight connection between the group structure implied by the relativity principle and the rules for the transformation of space-time coordinates~\cite{Ignatowsky:1910a, Ignatowsky:1910b, Ignatowsky:1911a, Ignatowsky:1911b}. Under suitable hypotheses, (explicitly: locality, linearity, isotropy), the relativity principle almost uniquely leads to either the Lorentz transformations or Galileo's transformations, and this result makes no \emph{a priori} appeal to the constancy of the speed of light. Over the last century this same result has been repeatedly rediscovered, expanded upon, and re-analyzed, with significant  pedagogical efforts being expended; see for instance~\cite{fr1, fr2, whitehead, pars, lalan, severi, terletskii, sussmann, Berzi:1969, Gorini:1970, Gorini:1971, Lugiato:1972, Berzi:1972, Lee-Kalotas, Levy-Leblond, Srivastava, rindler, jammer, torretti, Fock, gr-qc/0107091}. The relevance of von Ignatowsky's analysis for our current purposes comes from reversing the logic: If for whatever reason one wishes to speculate about a possible breakdown of Lorentz invariance at ultra-high energies, then as long as one continues to work within the framework of  classical local physics one is almost certain to be forced to abandon (or at the very least grossly modify) the relativity principle --- and in particular one will in general be forced to abandon the group structure for the set of transformations that connect space and time in one inertial frame to space and time in a different inertial frame. 

This is the basic theme of this article: \emph{Within the framework of local physics, what happens to inertial frames, and the transformations between inertial frames, if you do not have the relativity principle?} We shall see that quite a lot can still be said. Under suitable hypotheses, it is possible to argue that the (space and time) transformation rules  between inertial frames should at least be linear. (We shall subsequently have a few words to say about situations where these transformations might not be linear.) For linear transformations between inertial frames,  rather general formulae for the transformation of 3-velocities, and in particular the composition of 3-velocities, can then be derived.  

We shall see that in local linear theories with a preferred frame (an ``aether'') the set of transformations between inertial frames forms a \emph{groupoid}/\emph{pseudogroup}. (In this particular sub-branch of mathematics the  mathematical terminology is not 100\% settled.) This \emph{groupoid}/\emph{pseudogroup} structure is not just a mathematical curiosity, the distinction between a  \emph{groupoid}/\emph{pseudogroup} and a \emph{group} is essential to evading von Ignatowsky's 1910 argument. 

In reference~\cite{gr-qc/0107091} one of the current authors had, with collaborators, explored the possibility of, in certain circumstances, permitting ``faster than light'' signals while still retaining the relativity principle, (and locality and linearity, but not isotropy). In the current article we shall instead more radically explore what happens if it is the relativity principle that is sacrificed. 

Now, ``merely'' knowing how space and time transform is not sufficient to do anything beyond the most basic of kinematics. A key ingredient to understanding physics beyond kinematics is to have some notion of dynamics. To do so requires one (at an absolute minimum) to develop some sort of notion of energy and momentum --- thereby implying the ability to construct some sort of Lagrangian or Hamiltonian mechanics.  Specifically, to understand dynamics in Lorentz violating theories it is necessary to understand how energy and momentum transform, and to understand how the Lagrangian and Hamiltonian transform. This is more subtle than one might naively expect. In local theories with linear transformations between inertial frames, the generic situation is that energy and momentum transform in an \emph{affine} manner (that is, linear plus an inhomogeneous offset term). 

We shall show that it is possible, but not always desirable, to choose conventions and parameters in such a way as to force the offset to be zero --- in which case energy and momentum transform in a homogeneous linear manner. In fact, if this is done, then with our conventions the 4-component vector $P=(E,-\p^T)$, the 4-momentum, is a row vector. $P$ is an element of the vector space dual to the 4-component vector $X= (t,\x^T)^T$, the 4-position, which is a column vector. (We shall see that the offset term in the affine transformation is needed if one wishes to recover the usual naive form of Newtonian  mechanics in a suitable limit, but that there is a somewhat non-standard formulation of Newtonian mechanics in which energy-momentum can be made to transform linearly.) 

Finally, as a side effect of the general analysis, we focus specific attention on two particularly elegant and compelling models implementing a ``minimalist'' violation of Lorentz invariance. The first minimalist Lorentz-violating model confines all Lorentz violating physics to some suitable sub-sector of the particle physics spectrum (most typically taken to be the neutrino sector). The second minimalist Lorentz-violating model preserves as much as possible of standard Lorentz invariant physics, but the transformations additionally depend on one extra function, an arbitrary free function of absolute velocity.  Consequently,  when studying possible violations of Lorentz invariance, these two models in many ways serve as  examples of ``least-damage'' violations of Lorentz invariance.
Indeed, the considerations of this article will be essential to almost any form of violation of Lorentz invariance that respects locality and encodes ``preferred frame'' (aether frame) effects.

\section{Why violate Lorentz invariance?}

Why is there currently such significant interest in the possibility of broken Lorentz invariance? There are several reasons, based on a variety of considerations. There are purely theoretical speculations, there is the very practical and pragmatic need for a phenomenological framework within which to formulate empirical tests of Lorentz invariance, and there are even some rather \emph{tentative} experimental hints of the observation of violations of Lorentz invariance.

\subsection{Theoretical considerations} 

There have been numerous and long-standing theoretical suggestions to the effect that quantum gravity might eventually violate Lorentz invariance at ultra-high energies. For instance, there is the string-inspired theoretical framework for characterizing possible violations of Lorentz invariance developed by Kostelecky and collaborators~\cite{Colladay:1998fq, Kostelecky:1988zi, Kostelecky:2003fs, Kostelecky:2000mm, Kostelecky:2002hh, Kostelecky:2003cr, Kostelecky:1999mr, Kostelecky:2001mb, Bear:2000cd}. More recently, the Horava gravity framework~\cite{Horava:2009uw} naturally includes Lorentz violation~\cite{Visser:2009fg, Visser:2009ys, Sotiriou:2009bx, Sotiriou:2009gy, Weinfurtner:2010hz, Visser:2011mf}. The ``analogue spacetime'' programme also very naturally leads to models where Lorentz invariance is violated at one level or another~\cite{Barcelo:2005fc, Visser:1997ux, Broken, BEC}. There is also the flat-space non-gravity framework developed by Anselmi~\cite{Anselmi:2011zz, Anselmi:2011bp, Anselmi:2007zz, Anselmi:2011ae, Anselmi:2011ac, Anselmi:2010zh, Anselmi:2009ng, Anselmi:2009vz, Anselmi:2008bt, Anselmi:2008bs, Anselmi:2008bq, Anselmi:2008ry, Anselmi:2007ri}, where Lorentz invariance breaking is used to partially regulate QFT ultraviolet divergences. Further afield, Nielsen and collaborators have studied the renormalization group flow of Lorentz symmetry violating operators in generic QFTs, demonstrating that Lorentz invariance is often an infrared fixed point of a generic Lorentz violating QFT~\cite{NBI-HE-78-10, NBI-HE-82-42, NBI-HE-82-30, 187008, NBI-HE-82-9}. That is, there is an already vast literature regarding possible violations of Lorentz symmetry, and we have necessarily had to be rather selective in choosing citations.

\subsection{Phenomenological considerations}

 If one wishes to observationally test Lorentz invariance one needs some coherent framework to work in that at least allows one to formulate appropriate questions. Over the last decade significant progress along these lines has been made. See for instance work by Coleman and Glashow~\cite{Coleman:1997xq, Coleman:1998ti},  Amelino-Camelia, Ellis, Mavromatos, Nanoplous, and Sarkar~\cite{AmelinoCamelia:1997gz}, Gambini and Pullin~\cite{Gambini:1998it}, Kifune~\cite{Kifune:1999ex}, Aloisi, Blasi, Ghia, and Grillo~\cite{Aloisio:2000cm}, Amelino-Camelia and Piran~\cite{AmelinoCamelia:2000zs}, plus that by 
Jacobson, Liberati,  and Mattingly~\cite{Liberati:2001cr, Jacobson:2001tu,  Jacobson:2002hd, Mattingly:2002ba, Jacobson:2002ye, Jacobson:2003ty, Jacobson:2003bn, Jacobson:2004qt, Jacobson:2004rj, Jacobson:2005bg}, and especially the Living Review by Mattingly~\cite{Mattingly:2005re}. The net result of all these efforts is that we now have a considerable quantity of observational bounds, some of them very stringent observational bounds,  constraining the possibility of Lorentz symmetry breaking --- although it should perhaps be noted that these analyses are typically performed in the preferred (aether) frame. 

\subsection{Experimental considerations} 

The OPERA experiment has recently announced rather \emph{tentative} but statistically significant evidence for Lorentz symmetry violation in the form of ``faster than light'' neutrinos~\cite{Opera:2011}. (See also earlier even more tentative results from the MINOS collaboration~\cite{Minos:2007}.)
In the resulting firestorm, over 330 theoretical articles have been generated in some $6$ months.  
Notable contributions include~\cite{AmelinoCamelia:2011dx, AmelinoCamelia:2011bz, Giudice:2011mm, 
Cohen:2011hx, Dvali:2011mn, Alexandre:2011bu, Cacciapaglia:2011ax, Bi:2011nd, 
Klinkhamer:2011mf, Gubser:2011mp, 
Kehagias:2011cb, Wang:2011sz, Saridakis:2011eq, Winter:2011zf, Alexandre:2011kr, Klinkhamer:2011iz, Cowsik:2011wv, Maccione:2011fr, Dass:2011yj, Carmona:2011zg, arXiv:1111.6340}. 
We particularly wish to emphasise the importance of the experimental/observational bounds presented in the Cohen--Glashow~\cite{Cohen:2011hx} and Maccione--Liberati--Mattingly articles~\cite{Maccione:2011fr}. The consensus opinion now seems to be that the specific observations reported by OPERA were contaminated by significant experimental errors, and that their tentative observations are ruled out by newer results~\cite{ICARUS}, but the episode has nevertheless inspired significant interest in Lorentz-violating phenomenological models. While our own interest in these issues was in many ways OPERA-MINOS-inspired, we emphasise that our analysis and conclusions as presented in this article are in no way specifically OPERA-MINOS-dependent.


\subsection{DSR?} 

Regarding the possibility of working with DSR (doubly special relativity, distorted special relativity), it will soon become clear that DSR falls outside our framework. (For general background see~\cite{gr-qc/0012051, arXiv:1111.5643, gr-qc/0207049, Judes:2002bw, Liberati:2004ju}.)  For DSR-like models a  key issue is that after a decade of work on this topic, and despite significant ongoing efforts,  there is still no clear universally accepted consensus as to how space and time transform between inertial frames~\cite{Girelli:2004ue, Girelli:2005dc, Smolin:2010xa, Deriglazov:2004yr, Hossenfelder:2007fy, Aloisio:2004rd, Aloisio:2005rq, Schutzhold:2003yp, Grumiller:2003df, Ahluwalia:2002wf, Rembielinski:2002ic, Calmet:2010tx, Hossenfelder:2010tm, Hossenfelder:2009mu,  Hossenfelder:2006rr, Hossenfelder:2010xr, Hossenfelder:2010yp, Rovelli:2008cj, Hinterleitner:2004ny} --- there is not even any clear consensus on whether or not photon velocities are momentum-dependent in general DSR frameworks~\cite{Daszkiewicz:2003yr, Ghosh:2006cb}.  There are also suggestions to the effect that the ``D'' in DSR should be attributed to adopting a modified theory of measurment~\cite{Liberati:2004ju}.

However, there \emph{is} reasonable consensus that the energy-momentum transformations of DSR-like theories are generically of the form~\cite{Judes:2002bw}
\begin{equation}
P \to \bar P = f(L f^{-1}(P)),
\end{equation}
for $L$ an ordinary linear Lorentz transformation and $f(P)$ some nonlinear function on energy-momentum space. Since these energy-momentum transformations are \emph{not} affine, the considerations of this article imply that DSR-like theories (insofar as they are internally consistent), must at the very least exhibit other oddities --- such as a breakdown in locality, or a breakdown in linearity, (which we shall soon see implies a breakdown in the usual notion of inertial frame), or a breakdown of the existence of any notion of Hamiltonian/Lagrangian mechanics --- any of which would then undermine the very notions of energy and momentum used to define the DSR energy-momentum transformations in the first place.

\subsection{Relative locality?} 

Additionally, there have recently been some speculations (and some significant disagreements) in the literature regarding non-local models based on so-called ``relative locality''~\cite{AmelinoCamelia:2010qv, Hossenfelder:2010jn, AmelinoCamelia:2011yi, AmelinoCamelia:2011bm, AmelinoCamelia:2011pe, Hossenfelder:2012vk}. Currently, these models are still being developed and investigated. Certainly, they very explicitly fall outside the framework we are considering in this article. 

Roughly speaking, in relative locality models it is momentum space that is taken to be primary, with single-particle phase space being the tangent bundle to momentum space --- the various tangent spaces [indexed by the 4-momentum] then correspond to logically distinct spacetimes indexed by the 4-momentum of the particle being observed. More generally, in multi-particle contexts these relative locality models appear to generalize/modify the notion of inertial frame in such a way that  it  depends not only on the state of motion of the observer, but also on the collection of 4-momenta of the various objects being observed.

\subsection{Pragmatic considerations} 

There is an experimental adage to the effect that one should ``only adjust one parameter at a time'' --- the theorist's equivalent is that one should ``only adjust one theoretical assumption at a time''.  Controlled restraint in relaxing one's input assumptions  is essential if one is to isolate exactly \emph{which} theoretical aspect is critical to \emph{which} phenomenological result.  

There is a trade-off between generality and precision. A framework that is too abstract and flexible has difficulty making any precise or definite predictions.  Controlled restraint in relaxing one's input assumptions  is essential if one is to develop a pragmatically useful framework that is sufficiently well-defined to make definite statements that can in principle be confronted with empirical reality.

\subsection{Summary} 

Given this level of interest in the topic, we have feel that it is interesting, useful, and timely to perform a careful analysis of the general and very basic notion of inertial frames in the \emph{absence} of Lorentz invariance. We shall focus particularly on ``preferred frame''  (aether) versions of Lorentz symmetry breaking --- that is, we shall study inertial frames in the \emph{absence} of the relativity principle, but while retaining usual notions of local physics.

\section{General transformations between inertial frames}
\subsection{Definition of an inertial frame}

Essentially everyone would agree on this characterization of inertial frames:
\begin{itemize}
\item 
All inertial frames are in a state of constant, rectilinear motion with respect to one another; they are not accelerating (in the sense of proper acceleration that would be detected by an accelerometer). 
\item
In an inertial reference frame, the laws of mechanics take their simplest form.
\item
In an inertial frame, Newton's first law (the law of inertia) is satisfied: Any free motion has a constant magnitude and direction, implying a linear relationship between the space and time coordinates assigned to any free particle.  
\end{itemize}
If \emph{in addition} you accept the relativity principle, then:
\begin{itemize}
\item 
Physical laws take the same form in all inertial frames.
\end{itemize}
But, as we have argued above, for some purposes the relativity principle is overkill.
And that is the topic we will now explore.

An implicit assumption in this discussion is that the space and time coordinates are specified in terms of real numbers --- this is a purely pragmatic decision based on the very well-developed theories of differential and integral calculus (and ODEs and PDEs) which are available for the real number system, but which are much more limited for other mathematical frameworks. If we wish to investigate dynamics, or even non-trivial kinematics, we had better be able to at least differentiate (and anti-differentiate) --- which immediately forces us to focus attention on the real numbers.  (A much more radical proposal would be to discard all notions of differentiability and integrability completely, and to base all fundamental physics directly on finite-difference equations. There are real and significant costs to any such proposal, notably with regard to existence and uniqueness of solutions to finite-difference dynamical equations, and we will not further explore that particular option.)

\subsection{Argument for linearity}

Even in the absence of the relativity principle, we shall still want the transformations between inertial frames to be linear. 
By definition a freely moving particle, in an inertial frame,  is not accelerating 
\begin{equation}
{d^2 \x \over dt^2} = 0; \qquad \x(t) = \x_0 + \v_0 t.
\end{equation}
In any other inertial frame, the particle is again by definition not accelerating 
\begin{equation}
{d^2 \bar\x \over d\bar t^2} = 0; \qquad \bar\x(\bar t) = \bar\x_0 + \bar\v_0 \bar t.
\end{equation}
Whatever the transformation is between the two sets of time and space coordinates  $\{t,\x\}$ and $\{\bar t, \bar \x\}$, the transformation has to map straight lines into straight lines --- 
which forces the transformation to be, at the very worst, projective~\cite{rindler, Fock}. By additionally requiring that events in a bounded region map into a bounded region this is actually enough to force 
 the transformations to be linear~\cite{rindler, Fock}. If one is \emph{only} interested in implementing Newton's first law, then one could get away with using an arbitrary abstract vector space defined over an arbitrary abstract number field. Newton's first law would still require straight lines to map into straight lines, and so would naturally lead to projective transformations.  By considering bounded regions of a normed vector space it becomes natural to restrict attention to linear transformations between inertial frames. But as soon as one wishes to implement Newton's second law, one needs to be able to discuss non-zero variable acceleration, which requires some notion of differentiability, which naturally leads one to consider a vector space defined over the real numbers.

Writing the 4-position as 
\begin{equation}
X = \left(\begin{array}{c} t \\ \x \end{array}\right),
\end{equation}
we want
\begin{equation}
X \to \bar X = M \,X.
\end{equation}
Note that we adopt conventions where both 3-vectors $\x$ and 4-vectors $X$ are column vectors. This minimizes the number of special case notational fiddles we have to adopt later on in the discussion.

Note that the primary physics input is the  \emph{observation} that inertial frames exist, and from extremely basic notions of kinematics this is enough to argue for linearity. If one additionally (as we shall see later in the article) wants to develop some notion of Lagrangian/Hamiltonian dynamics, then the observation that free inertial particles exist, coupled with Noether's theorem, can be used to argue for 
the homogeneity of space and time. Some authors prefer to start from homogeneity, and thereby deduce linearity. 
There are minor technical issues, but for all practical purposes spacetime homogeneity implies and is implied by linearity of the transformations between inertial frames. 

We emphasise how basic and fundamental the argument for linearity is --- if the transformation law between inertial frames is not linear, then it is really the whole notion of inertial frame that is being undermined. If, as is often (but not always) done, one interprets DSR as implying non-linear space and time transformations, then one is forced to abandon the very notion of inertial frame. With this interpretation DSR-based models generalize the relativity principle, but sacrifice inertial frames, DSR-based models then become ``relativity without inertial frames'',  whereas the whole thrust of this current article is ``inertial frames without relativity''.

\subsection{General representation of inertial transformations}

Taking linearity as given:
\begin{itemize}

\item 
In the special case of Newtonian physics (Galilean relativity) we have
\begin{equation}
M = \left[\begin{array}{c|c} \vphantom{\Big|}1 & \0^T \\ \hline \vphantom{\Big|}-\v & I \end{array} \right].
\end{equation}
\item
In the case of Einstein physics (special relativity) we have
\begin{equation}
M = \left[\begin{array}{c|c} \vphantom{\Big|}\gamma & -\gamma \v^T/c^2 \\ \hline 
\vphantom{\Big|}-\gamma \v &  \gamma\n \n^T  + [I-\n \n^T] \end{array} \right],
\end{equation}
with $\v= v \n$ and $\gamma = (1-v^2/c^2)^{-1/2}$. 
One can also write this as
\begin{equation}
M = \left[\begin{array}{c|c} \vphantom{\Big|}\gamma & -\gamma \v^T/c^2 \\ \hline 
\vphantom{\Big|}-\gamma \v &  \gamma\n \otimes \n  + [I-\n\otimes \n] \end{array} \right].
\end{equation}
\item
Carroll kinematics (``Alice in wonderland kinematics'') is a rarely encountered and somewhat unphysical limit of the Lorentz group where one takes $c\to0$ and $v\to0$ while keeping the ``slowness'' $u=v/c^2$ fixed. The resulting transformations
\begin{equation}
M = \left[\begin{array}{c|c} \vphantom{\Big|}1 & -\u^T \\ \hline 
\vphantom{\Big|} \0 &  I \end{array} \right]
\end{equation}
correspond to~\cite{Carroll1, dyson, hep-th/0410086, Carroll2, Carroll3}:
\begin{eqnarray}
t \to \bar t = t - \u \cdot \x; \qquad \x \to\bar \x = \x.
\end{eqnarray}
We will have very little to say concerning this particular option.
\end{itemize}
In the general case, we only know that $M$ is \emph{some} matrix which we can, without loss of generality, write in the form
\begin{equation}
M = \left[\begin{array}{c|c} \vphantom{\Big|}\gamma & -\u^T \\ \hline 
\vphantom{\Big|}-\w &  \Sigma \end{array} \right];
\qquad
M^{-1} = \left[
\begin{array}{c|c} \vphantom{\Big|} 
(\gamma-\u^T \Sigma^{-1} \w)^{-1} & (\u^T/\gamma) (\Sigma - \w\u^T/\gamma)^{-1}\\ \hline 
\vphantom{\Big|}
(\gamma-\u^T \Sigma^{-1} \w)^{-1}  \Sigma^{-1} \w &  (\Sigma - \w\u^T/\gamma)^{-1}
\end{array} 
\right].
\end{equation}
Specifically, we are \emph{not} at this stage assuming any notion of isotropy.   Note that the object $\w \u^T= \w\otimes\u$ is a $3\times 3$ matrix, while $\u^T \w = \u\cdot \w$ is a scalar. By replacing $\u\to\gamma\u$, we could also choose to write this in the completely equivalent form
\begin{equation}
M = \left[\begin{array}{c|c} \vphantom{\Big|}\gamma & -\gamma\u^T \\ \hline 
\vphantom{\Big|}-\w &  \Sigma \end{array} \right];
\qquad
M^{-1} = \left[
\begin{array}{c|c} \vphantom{\Big|} 
\gamma^{-1} (1-\u^T \Sigma^{-1} \w)^{-1} & \u^T (\Sigma - \w\u^T)^{-1}\\ \hline 
\vphantom{\Big|}
\gamma^{-1} (1-\u^T \Sigma^{-1} \w)^{-1}  \Sigma^{-1} \w &  (\Sigma - \w\u^T)^{-1}
\end{array} 
\right].
\end{equation}
Alternatively, by now replacing $\w\to\Sigma\,\w$, 
\begin{equation}
M = \left[\begin{array}{c|c} \vphantom{\Big|}\gamma & -\gamma\u^T \\ \hline 
\vphantom{\Big|}-\Sigma \w &  \Sigma \end{array} \right];
\qquad
M^{-1} = \left[
\begin{array}{c|c} \vphantom{\Big|} 
\gamma^{-1} (1-\u^T  \w)^{-1} & \u^T (I - \w\u^T)^{-1} \Sigma^{-1}\\ \hline 
\vphantom{\Big|}
\gamma^{-1} (1-\u^T \w)^{-1}  \w &  (I - \w\u^T)^{-1} \Sigma^{-1}
\end{array} 
\right].
\end{equation}
Any one of these three ways of parameterizing the $4\times4$ matrix $M$ is completely equivalent and mathematically acceptable, and which one we adopt is simply a matter of taste. 
(It is easy to explicitly carry out the matrix multiplications to verify that $M M^{-1}=I$.)

\subsection{Aether frame and moving frame}

Let us now distinguish two frames, the aether frame (the preferred rest frame) $F$ with coordinates $X$, and the moving frame $\bar F$ with coordinates $\bar X$. Then for definiteness we will choose $M$ to map from the aether frame to the moving frame, and $M^{-1}$ to map from the moving frame to the aether frame, so that
\begin{equation}
\bar X = M \,X; \qquad 
X = M^{-1} \,\bar X. 
\end{equation}
(Choosing which of the frames is the aether, and which is moving,  is merely a matter of convention.) 
We shall now rename things slightly and shall henceforth adopt the specific convention and nomenclature that:
\begin{equation}
M = \left[\begin{array}{c|c} \vphantom{\Big|}\gamma & -\gamma\u^T \\ \hline 
\vphantom{\Big|}-\Sigma \v &  \Sigma \end{array} \right];
\qquad
M^{-1} = \left[
\begin{array}{c|c} \vphantom{\Big|} 
\gamma^{-1} (1-\u^T  \v)^{-1} & \u^T (I - \v\u^T)^{-1} \Sigma^{-1}\\ \hline 
\vphantom{\Big|}
\gamma^{-1} (1-\u^T \v)^{-1}  \v &  (I - \v\u^T)^{-1} \Sigma^{-1}
\end{array} 
\right].
\end{equation}
Note that with these conventions both $\gamma$ and $\Sigma$ are dimensionless, while $\v$ has the dimensions of velocity, and $\u$ has dimensions of ``slowness'' = 1/(velocity). 
It is again easy to verify that $M M^{-1}=I$.
It is also useful to note (see appendix \ref{A:identities}) 
\begin{equation}
\u^T (I - \v \u^T)^{-1} = (1-\u^T \v)^{-1} \u^T = (1-\u \cdot \v)^{-1} \u^T,
\end{equation}
and so write
\begin{equation}
M = \left[\begin{array}{c|c} \vphantom{\Big|}\gamma & -\gamma\u^T \\ \hline 
\vphantom{\Big|}-\Sigma \v &  \Sigma \end{array} \right];
\qquad
M^{-1} = \left[
\begin{array}{c|c} \vphantom{\Big|} 
\gamma^{-1} (1-\u \cdot\v)^{-1} & (1 - \u\cdot\v)^{-1} \, \u^T \Sigma^{-1}\\ \hline 
\vphantom{\Big|}
\gamma^{-1} (1-\u \cdot\v)^{-1}  \v &  (I - \v\otimes\u)^{-1} \Sigma^{-1}
\end{array} 
\right].
\end{equation}
Note that there is a kinematic singularity if $\u\cdot\v=1$; in the particular case of special relativity this would correspond to an infinite boost to a frame traveling at lightspeed. But the possible occurrence of these kinematic singularities is a much more general phenomenon, and is not limited to special relativity. Indeed, since with our conventions the matrix $M$ factorizes
\begin{equation}
M = 
\left[\begin{array}{c|c} \vphantom{\Big|}\gamma & \0^T \\ \hline 
\vphantom{\Big|} \0 &  \Sigma \end{array} \right] 
\;
\left[\begin{array}{c|c} \vphantom{\Big|} 1&  - \u^T \\ \hline 
\vphantom{\Big|}- \v &  I \end{array} \right],
\end{equation}
we see that
\begin{equation}
\det(M) = \gamma \det(\Sigma) \;\det\!\!\left[\begin{array}{c|c} \vphantom{\Big|} 1&  - \u^T \\ \hline 
\vphantom{\Big|}- \v &  I \end{array} \right] = \gamma \det(\Sigma) \; [1-\u\cdot \v].
\end{equation}
So the existence of the kinematic singularity is equivalent to the non-invertibility of the transformation matrix, a possibility that should (generically) be excluded on physical grounds.

An object that is at rest in the moving frame follows the worldline
\begin{equation}
\bar X =  \left({\bar t\atop\0}\right),
\end{equation}
which in the aether frame coordinates maps into
\begin{equation}
X =  M^{-1} \,\bar X = \bar t \; \gamma^{-1} (1-\u \cdot \v)^{-1}  \left({1\atop\v}\right). 
\end{equation}
This implies $\x = \v t$. 
That is,  with these conventions the moving frame has 3-velocity $\v_\mathrm{moving}=\v$ as viewed by the aether, and this is our physical interpretation of the parameter $\v$ appearing in the matrix $M$. 
But what about the aether frame as seen by the moving frame?  An object at rest in the aether frame follows the worldline
\begin{equation}
X = \left({t\atop\0}\right), 
\end{equation}
which in the moving frame coordinates maps into
\begin{equation}
\bar X = M \, X =  t \;\left(\begin{array}{c}  \gamma \\ - \Sigma \v\end{array}\right).
\end{equation}
This implies $\bar\x = -(\Sigma\v/\gamma) \bar t$.
That is, as viewed in the moving frame, the aether is moving with 3-velocity
\begin{equation}
\v_\mathrm{aether} = -{\Sigma\, \v \over\gamma}. 
\end{equation}
 Note that $\v_\mathrm{moving}$ and $\v_\mathrm{aether}$ are generally \emph{not} equal-but-opposite velocities. In fact, without additional assumptions, in the general case they need not even be collinear. 

\subsection{Transformation of 3-velocity}

From $\bar X = M\, X$ we have $d\bar X = M \,d X$, whence with the conventions adopted above we see
\begin{equation}
d\bar t = \gamma(dt - \u\cdot d\x ); \qquad d\bar\x = \Sigma(d\x-\v dt),
\end{equation}
so that
\begin{equation}
\dot{\bar\x} \equiv {d\bar\x\over d\bar t} =  {\Sigma(\dot\x-\v)\over\gamma(1-\u\cdot\dot\x)}\; .
\end{equation}
This is the general combination of velocities rule. (One can easily see that it is a natural generalization of the usual special relativistic combination of velocities, with current conventions being chosen to make this transformation as simple as possible).  Note in particular that an object at rest in the aether frame, with $\dot\x=\0$, moves at 3-velocity $-\Sigma\v/\gamma$ in the moving frame, while an object at rest in the moving frame, with $\dot{\bar\x}=\0$, moves at 3-velocity $\v$ in the aether frame.

Similarly, from $d X = M^{-1} \, d\bar X$ we have
\begin{equation}
d t =  \gamma^{-1} (1-\u\cdot  \v)^{-1} d\bar t + (1 - \u\cdot \v)^{-1} \u^T \Sigma^{-1}  d\bar \x,
\end{equation}
and
\begin{equation}
 d\x =   (I - \v\otimes\u)^{-1} \Sigma^{-1} d\bar\x + \gamma^{-1} (1-\u\cdot \v)^{-1}  \v d\bar t,
\end{equation}
so that
\begin{equation}
\dot\x \equiv {d\x\over dt} =   {
(I - \v\otimes\u)^{-1} \Sigma^{-1} \dot{\bar\x} + \gamma^{-1} (1-\u \cdot\v)^{-1}  \v
\over
 \gamma^{-1} (1-\u \cdot \v)^{-1}  + (1 - \u\cdot\v)^{-1} \u^T \Sigma^{-1}  \dot{\bar \x}
}\;.
\end{equation}
We can simplify this to obtain
\begin{equation}
\dot\x =   {
\gamma (1-\u \cdot\v) (I - \v\otimes\u)^{-1} \Sigma^{-1} \dot{\bar\x} +  \v
\over
 1 + \gamma\u^T \Sigma^{-1}  \dot{\bar \x}
}.
\end{equation}
But (see appendix \ref{A:identities})
\begin{equation}
(1- \u\cdot \v)(I - \v\otimes\u)^{-1} 
=  (1- \u\cdot \v) I + \v\otimes \u,
\end{equation}
and so
\begin{equation}
\dot\x =   {
\gamma [(1- \u\cdot \v) I + \v\otimes \u] \, \Sigma^{-1}   \dot{\bar\x} +  \v
\over
 1 + \gamma\u^T \Sigma^{-1}  \dot{\bar \x}. 
}.
\end{equation}
Finally
\begin{equation}
\dot\x =   {
\gamma (1- \u\cdot \v) \, \Sigma^{-1}   \dot{\bar\x} 
\over
 1 + \gamma\u^T \Sigma^{-1}  \dot{\bar \x} }  + \v. 
\end{equation}
This last formula seems to be the best one can do. Attempting to change conventions to simplify \emph{this} particular formula leads to problems elsewhere.
Note in particular that something at rest in the moving frame, so that $\dot{\bar\x}=0$, moves at velocity $\v$ in the aether frame.

\subsection{Groupoid/pseudogroup structure}

Physically the matrix $M$ need not be a function of $\v$ only --- it can also depend on the orientation of the moving inertial frame with respect to the preferred frame, and worse the quantities $\gamma$, $\u$, and $\Sigma$, are (potentially) free parameters in their own right. It is useful to write $M(\bar F)$, to emphasise that the matrix $M(\bar F)$ is potentially a function of \emph{all} the parameters characterizing the moving inertial frame $\bar F$.  (We could furthermore write the various pieces of $M(\bar F)$ as $\gamma(\bar F)$, $\v(\bar F)$, $\u(\bar F)$, and $\Sigma(\bar F)$; while technically more correct, this is so clumsy as to be impracticable, and the frame dependence of these quantities will always be implicitly understood.) 

In addition, one should keep in mind that in general the transformation matrices $M(\bar F)$ could also depend on the internal structure of the particular type of rulers and clocks one is using; it is only for situations of very high symmetry --- essentially amounting to adoption of the relativity principle --- that the notion of time and distance can be abstracted to have a meaning that is independent of the internal structure of one's choice of clocks and rulers.

Note that in general the set $\{ M(\bar F) \}$, (where $M(\bar F)$ is the transformation matrix from the aether inertial frame to the moving inertial frame $\bar F$),  need not form a group; and  similarly the set $\{ M^{-1}(\bar F) \}$ need not form a group.
There is no need for these sets to be closed under matrix multiplication. Nor are these sets generally closed under matrix inversion. 
There does not seem to be any specialized mathematical terminology for such objects --- they are not semigroups, they are not groupoids, they are not pseudogroups,  they are not monoids, they are not cosets, they are not magmas; they are just sets of matrices.

However to transform from one arbitrary inertial frame $F_1$ to another arbitrary inertial frame $F_2$, the appropriate transformation is
\begin{equation}
M(F_2,F_1) = M(F_2) \, M(F_1)^{-1}.
\end{equation}
The set $\{ M(F_2,F_1)\}  = \{ M(F_2)\, M(F_1)^{-1} \} $ certainly forms a \emph{groupoid}/\emph{pseudogroup}, in the sense that the set is closed under the restricted set of compositions (so-called \emph{partial-products}) of the form
\begin{equation}
 M(F_3,F_2)  M(F_2,F_1)  = M(F_3,F_1).
\end{equation}
(The relevant mathematical terminology is not 100\% standardized, and different sources prefer to call this mathematical structure either a \emph{groupoid} or a \emph{pseudogroup}.)
Note that $M(F,F)=I$, so an identity certainly exists, and that $M(F_2,F_1)^{-1} = M(F_1,F_2)$ so that inverses also exist. Associativity is automatic because matrix multiplication is associative.  
In general this is the most you can say.
The technical difference between a group and a \emph{groupoid}/\emph{pseudogroup} is in this context \emph{extremely important}. 
It is this technical mathematical distinction that ultimately allows us to side-step the usual von Ignatowsky theorems (and their variants) that under normal circumstances lead almost uniquely to the Lorentz group or the Galileo group --- the physics reason  for this extra generality is because while we have assumed linearity of the transformations we have \emph{not} assumed the relativity principle. (Nor have we at this stage assumed isotropy, but that is not important for this particular issue.) 

The key physics point is that without the relativity principle $M(F_1,F_2)$ need not depend merely on the relative velocities between the frames $F_1$ and $F_2$, but is instead allowed to depend on the separate absolute velocities of the frames $F_1$ and $F_2$.

\section{Transformations of energy and momentum}

Defining energy and momentum, as opposed to purely kinematical notions of velocity and position, requires at least some notion of dynamics. Pick some arbitrary but fixed inertial frame. To study dynamics in that frame, one should at a minimum be able to formulate all three of Newton's laws, and one should at a minimum be able to formulate notions of energy and momentum. Since both Hamiltonian and Lagrangian dynamics are essentially just a reformulation of Newton's three laws, and implicitly of the notions of energy and momentum, one should in each individual inertial frame be able to develop \emph{some} version of Hamiltonian/Lagrangian dynamics.  (Furthermore, we note that any attempt at developing usual notions of quantum physics requires one to first develop Hamiltonian/Lagrangian dynamics --- either to insert into the path integral formulation, or to serve as the basis for a quantum Hamiltonian underlying the Schroedinger equation.) The natural question then arises as to how the Hamiltonian/Lagrangian dynamics in different inertial frames are related to each other.\footnote{\emph{Some} interesting questions can nevertheless be dealt with by working within a fixed but arbitrary inertial frame and not worrying about the transformation rules. For instance, scattering and decay thresholds in Lorentz violating theories can usefully be dealt with in such a manner, see for instance~\cite{Coleman:1997xq, Coleman:1998ti, Liberati:2001cr, Jacobson:2001tu, Jacobson:2002hd, Mattingly:2002ba, Jacobson:2002ye, Jacobson:2003ty, Jacobson:2003bn, Jacobson:2004qt, Jacobson:2004rj, Jacobson:2005bg, Mattingly:2005re, Cohen:2011hx, Maccione:2011fr}. This has led us to spin off such considerations into a separate article~\cite{arXiv:1111.6340}.} Understanding this will tell us how energy and momentum in different inertial frames are related to each other.

\subsection{Defining energy and momentum}

Ignoring interactions for now, in view of the homogeneity of spacetime we shall assume that each particle has associated with it \emph{some} Lagrangian $L(\dot\x)$ which leads to a momentum $\p(\dot\x) = \partial L/\partial \dot\x$, and hence to a Hamiltonian $H(\p)$, which we shall typically just write as $E(\p)$.  Because of space-time homogeneity and the Hamiltonian/Lagrangian framework, Noether's theorem implies energy and momentum conservation:
\begin{equation}
\sum_\mathrm{in} E_i=\sum_\mathrm{out} E_i; \qquad\qquad \sum_\mathrm{in} \p_i=\sum_\mathrm{out} \p_i.
\end{equation}
Now the inertial equations $\ddot\x=0$ will be satisfied for any arbitrary $L(\dot\x)$.  (Note the absence of any explicit $t$ or $\x$ dependence.) To operationally determine a specific $L(\dot\x)$ that can usefully characterize a specific particle, you will have to perform a large number of collisions at various 3-velocities, compare input and output states, and data-fit to extract suitable $E_i(\dot\x)$ and $\p_i(\dot\x)$ corresponding to the various particles in your universe of discourse. 
Once this is done you can build a model for the $L_i(\dot\x)$  using
\begin{equation}
L_i(\dot\x) =  L_i(\0) + \int_{\0}^{\dot\x}   \p(\dot{\tilde\x})\cdot d\dot{\tilde\x}.
\end{equation}
Note that, in modelling the $\p_i(\dot\x)$, one would have to take into account the consistency condition $\nabla_{\dot\x}\times\p_i(\dot\x)$ required for this construction to be path independent in velocity space.  

But even after one has done this, the construction \emph{cannot be unique} --- for any set of \emph{constants} $\epsilon_i$ and $\bpi_i$ such that
$\sum_\mathrm{in} \epsilon_i=\sum_\mathrm{out} \epsilon_i$ and $\sum_\mathrm{in} \bpi_i=\sum_\mathrm{out} \bpi_i$ we see that the assignments
\begin{equation}
E_i \leftrightarrow E_i + \epsilon_i ;   \qquad \p_i \leftrightarrow \p_i + \bpi_i
\end{equation}
are physically indistinguishable. But that means the Lagrangians
\begin{equation}
L_i(\dot\x) \leftrightarrow L_i(\dot\x) - \epsilon_i + \bpi_i \cdot \dot\x
\end{equation}
are physically indistinguishable.  In terms of the action this means
\begin{equation}
S_i = \int L_i(\dot\x) dt \leftrightarrow  S_i = \int  L_i(\dot\x) dt - \epsilon_i (t_F-t_I) + \bpi_i \cdot (\x_F-\x_I)
\end{equation}
are physically indistinguishable --- which is physically and mathematically obvious in view of the fact that the two actions differ only by boundary terms. This intrinsic ambiguity in the definition of energy and momentum will (perhaps unfortunately) turn out to be important. One could try to resolve these ambiguities in a number of different ways:
\begin{itemize}
\item 
For instance, the ambiguity in momentum could be fixed by setting the momentum at zero velocity to be zero: $\p(\dot\x=\0)=\0$. 
Sometimes this works well, sometimes it does not.
\item 
The ambiguity in energy is equivalent to an ambiguity in rest energy $E(\dot\x=\0)$; attempting to set the rest energy to zero is often severely problematic. 
\end{itemize}
In general it is best to keep this freedom available in the calculation as long as possible.

\subsection{Affine versus linear transformations}

What can we now say about energy and momentum, and their transformation properties, using only linearity of the transformations between inertial frames? (Recall that we very specifically do not assume isotropy or any form of the relativity principle.) 

Consider a single particle, but multiple inertial frames. 
To even begin to talk about energy and momentum, in each frame one must be able to set up a suitable Lagrangian and Hamiltonian, and there should be some as yet unspecified relationship between the Lagrangians and Hamiltonians in these distinct inertial frames. Furthermore, extrema of the action as calculated in one inertial frame must coincide with extrema of the action calculated in any other inertial frame. 

That is, in complete generality we should demand that for any two inertial frames the action calculated in these frames should be equal up to boundary terms, and in each individual frame we know the action is ambiguous up to boundary terms. In view of the groupoid structure of the transformations between inertial frames there is no loss of generality in considering one moving frame $\bar F$ plus the aether frame $F$ for which we can write
\begin{equation}
\int \bar L \, d\bar t + \hbox{(boundary terms)} = \int L \, dt  + \hbox{(boundary terms)}.
\end{equation}
In view of our previous discussion this implies
\begin{equation}
\int \left\{ \bar L - \bar\epsilon+ \bar\bpi\cdot(d\bar\x/d\bar t)\right\} d \bar t
=
\int \left\{ L - \epsilon+ \bpi\cdot(d\x/dt)\right\} d t.
\end{equation}
But since $L = -(E-\p\cdot\dot\x)$  this implies
\begin{equation}
\int \left\{ (\bar E + \bar\epsilon)- (\bar\p+\bar\bpi)\cdot(d\bar\x/d\bar t)\right\} d \bar t
=
\int \left\{ (E+ \epsilon)-(\p +\bpi)\cdot(d\x/dt)\right\} d t.
\end{equation}
Therefore
\begin{equation}
 (\bar E + \bar\epsilon)\,d\bar t - (\bar\p+\bar\bpi)\cdot d\bar \x
=
(E+ \epsilon)\,dt -(\p +\bpi)\cdot d\x.
\end{equation}
So now in complete generality we have
\begin{equation}
(\bar E+\bar\epsilon, -\bar\p^T-\bar\bpi^T)\left({d\bar t\atop d\bar \x}\right)
= (E+\epsilon,-\p^T-\bpi^T)\left({dt\atop d\x}\right).
\end{equation}
But
\begin{equation}
\left({d\bar t\atop d\bar \x}\right) = M \, \left({dt\atop d\x}\right),
\end{equation}
therefore implying both
\begin{equation}
(E+\epsilon,-\p^T-\bpi^T) =  (\bar E+\bar\epsilon, -\bar \p^T-\bar \bpi^T) \, M,
\end{equation}
and 
\begin{equation}
(\bar E+\bar\epsilon, -\bar \p^T-\bar\bpi^T) =   (E+\epsilon,-\p^T-\bpi^T)\, M^{-1}.
\end{equation}
These are \emph{affine} transformation laws for energy and momentum, (that is, linear plus an inhomogeneous offset), with the affine piece only depending on the intrinsic ambiguities $\left(\epsilon, -\bpi^T\right)$ and  $\left(\bar\epsilon, -\bar\bpi^T\right)$  in the energy and momentum.
Note that $P=(E,-\p^T)$ transforms in the dual space [that is, dual to 4-position $X=(t,\x^T)^T\,$].
To be explicit about this
\begin{equation}
E \to \bar E =      {E-\p^T\v\over \gamma(1-\u^T\v)} + {\epsilon-\bpi^T \v\over \gamma(1-\u^T\v)} -\bar\epsilon,
\end{equation}
and
\begin{equation}
\p \to \bar \p =    (\Sigma^{-1})^T (I-\u\v^T)^{-1} (\p-E\u)+  (\Sigma^{-1})^T(I-\u\v^T)^{-1}(\bpi-\epsilon\u) -\bar\bpi.
\end{equation}
In terms of dot and tensor products we can rewrite this as
\begin{equation}
E \to \bar E =   {E-\p\cdot \v\over \gamma(1-\u\cdot \v)} + {\epsilon-\bpi \cdot\v\over \gamma(1-\u\cdot\v)} -\bar\epsilon,
\end{equation}
and
\begin{equation}
\p \to \bar \p =  (\Sigma^{-1})^T (I-\u\otimes\v)^{-1}  (\p-E\u)+ (\Sigma^{-1})^T (I-\u\otimes\v)^{-1} (\bpi-\epsilon\u) -\bar\bpi. 
\end{equation}
(One can now begin to see how the Lorentz and Galilean transformations might emerge as special cases of this very general result.)
The inverse transformations are somewhat simpler
\begin{equation}
\bar E \to E =   \gamma\bar E + \bar \p^T \Sigma \v  + \gamma\bar \epsilon + \bar \bpi^T \Sigma \v  -\epsilon,
\end{equation}
and
\begin{equation}
\bar\p \to \p = \gamma\bar E \u + \Sigma^T \bar \p  + \gamma\bar \epsilon \u + \Sigma^T \bar \bpi  -\bpi.
\end{equation}

Suppose we now consider the \emph{same} particle at two different 3-velocities, but working with the same two inertial frames $F$ and $\bar F$; then in terms of energy and momentum \emph{differences}, we can write a \emph{homogeneous} linear transformation law of the form
\begin{equation}
\left([\bar E_1-\bar E_2], -[\bar \p_1-\bar \p_2]^T\right) =   \left([E_1-E_2],-[\p_1-\p_2]^T\right) \;M^{-1}. 
\end{equation}
That is:
\begin{equation}
\Delta E \to \Delta \bar E =     {\Delta E- \Delta\p\cdot \v\over \gamma(1-\u\cdot \v)},
\end{equation}
and
\begin{equation}
\Delta\p \to \Delta\bar \p =   (\Sigma^{-1})^T (I-\u\otimes\v)^{-1}(\Delta\p-\Delta E\u).
\end{equation}
We need to compare the \emph{same} particle at two different velocities, since otherwise there is no particular reason for the $\left(\epsilon, -\bpi^T\right)$ and  $\left(\bar\epsilon, -\bar\bpi^T\right)$  to be the same for the two situations. 
Note that for two otherwise identical particles one could in principle choose differing values for the parameters  $\left(\epsilon, -\bpi^T\right)$ and  $\left(\bar\epsilon, -\bar\bpi^T\right)$, thereby making them distinguishable.   This does not appear to be what happens in our universe, so we shall assume that  the quantities $\left(\epsilon, -\bpi^T\right)$ and  $\left(\bar\epsilon, -\bar\bpi^T\right)$ , while they might depend on the inertial frame one is working in, are at least universal for any particular particle species. 

Note that in terms of energy-momentum differences the inverse transformations are
\begin{equation}
\Delta \bar E \to \Delta E =    \gamma\Delta \bar E + \Delta \bar \p^T \Sigma \v,
\end{equation}
and 
\begin{equation}
\Delta \bar\p \to \Delta \p =   \gamma\Delta \bar E \u + \Sigma^T \, \Delta \bar \p. 
\end{equation}
As a consistency check on the general formalism we can readily verify that these energy-momentum transformation laws are compatible with, and permit us to recover, the purely kinematical velocity combination rules. See appendix \ref{A:consistency} for details.

\subsection{Summary}

For each individual particle species we have
\begin{equation}
E \to \bar E =    {E-\p\cdot \v\over \gamma(1-\u\cdot \v)} + {\epsilon-\bpi \cdot\v\over \gamma(1-\u\cdot\v)}  -\bar\epsilon,
\end{equation}
and
\begin{equation}
\p \to \bar \p =  (\Sigma^{-1})^T  (I-\u\otimes\v)^{-1} (\p-E\u) +(\Sigma^{-1})^T  (I-\u\otimes\v)^{-1}(\bpi-\epsilon\u) -\bar\bpi, 
\end{equation}
while the inverse transformations are
\begin{equation}
\bar E \to E =   -\epsilon+ \gamma\bar E + \bar \p\cdot( \Sigma \v )+ \gamma\bar\epsilon + \bar \bpi \cdot(\Sigma \v), 
\end{equation}
and
\begin{equation}
\bar\p \to \p =  -\bpi +  \gamma\bar E \u + \Sigma^T \bar \p  +   \gamma\bar\epsilon \u + \Sigma^T \bar \bpi.
\end{equation}
We have a certain amount of freedom to choose $\epsilon$ and $\bpi$, and $\bar\epsilon$ and $\bar\bpi$. 
One obvious choice would be to always make the transformation laws linear; however as we shall soon see this is not always the best thing to do.

\section{Examples}

Let us now consider the very standard cases of Galilean invariance and Lorentz invariance, comparing affine and linear transformation laws for energy-momentum.

\subsection{Galileo group (affine version)}

For Galilean kinematics we have
\begin{equation}
M = \left[\begin{array}{c|c} \vphantom{\Big|}1 & \0^T \\ \hline \vphantom{\Big|}-\v & I \end{array} \right],
\end{equation}
so
\begin{equation}
\bar t = t; \qquad \bar\x = \x-\v t; \qquad \dot{\bar \x} = \dot\x - \v.
\end{equation}
Now \emph{one} natural choice is to choose the particularly simple and standard Lagrangians
\begin{equation}
L = {1\over2} m ||\dot\x||^2; \qquad \bar L = {1\over2} m ||\dot{\bar\x}||^2.
\end{equation}
(We shall soon see that there are also other choices one can make.)
Then
\begin{equation}
\bar L =  {1\over2} m ||\dot{\bar\x}||^2 =  {1\over2} m ||\dot{\x}-\v||^2 = 
 {1\over2} m ||\dot{\x}||^2 - m \v\cdot\dot\x + {1\over2} m ||{\v}||^2.
\end{equation}
That is
\begin{equation}
\bar L = L + {1\over2} m ||{\v}||^2 - m \v\cdot\dot\x.
\end{equation}
Now note
\begin{equation}
\bar\p = m\dot \x-m \v = m\dot{\bar\x}; \qquad  \bar H = \bar\p\cdot\dot{\bar\x}-\bar L = {||\bar\p||^2\over2m}.
\end{equation}
So working explicitly, with these particular conventions, we have \emph{affine} transformations for energy-momentum:
\begin{equation}
\bar E = E - \p\cdot \v + {1\over2} m ||\v||^2;
\qquad 
\bar\p =\p - m\v.
\end{equation}
In contrast, working directly from the general transformation laws derived above, and taking $\gamma=1$, $\u=0$, and $\Sigma=I$,
 we have
\begin{equation}
E \to \bar E =    -\bar\epsilon +  E-\p\cdot \v  + [\epsilon-\bpi\cdot \v] =  E-\p\cdot \v  + {1\over2} m ||\v||^2,
\end{equation}
and
\begin{equation}
\p \to \bar \p = -\bar\bpi + \p + \bpi = \p - m\v, 
\end{equation}
from which we deduce that this particular way of implementing Galilean mechanics corresponds to the choices
\begin{equation}
\bar\epsilon =  -{1\over2} m ||{\v}||^2; \qquad \bar\bpi = m \v; \qquad\qquad  \epsilon=0; \qquad \bpi = 0.
\end{equation}
(Remember that by convention $\bar F$ is the moving frame while $F$ is the ``aether'' frame. Note that it is the quantities $\{\bar\epsilon,\bar\bpi\}$ associated with the moving frame that are non-zero, and that these quantities depend on the velocity $\v$ of the moving frame.)
The inverse transformations are
\begin{equation}
\bar E \to E =   -\epsilon+ \bar E + \bar \p \cdot \v +\bar\epsilon +\bar\bpi \cdot\v
= \bar E + \bar\p\cdot\v + {1\over2} m ||\v||^2,
\end{equation}
and
\begin{equation}
\bar\p \to \p =   -\bpi + \bar \p +\bar\bpi = \bar\p + m\v.
\end{equation}
This is the ``usual'' way of doing Galilean dynamics, which unavoidably leads to \emph{affine} transformations for energy and momentum.

A somewhat subtle message to be taken from the discussion is this: Since affine transformations arise so naturally in this extremely straightforward setting, it seems unlikely that the affine features of the energy-momentum transformations could always be completely eliminated in more general settings.

\subsection{Lorentz group (linear version)}

In this case the Lorentz transformations are
\begin{equation}
M = \left[\begin{array}{c|c} \vphantom{\Big|}\gamma & -\gamma \v^T/c^2 \\ \hline 
\vphantom{\Big|}-\gamma \v &  \gamma\n \otimes \n  + [I-\n \otimes \n] \end{array} \right]
\end{equation}
with $\v= v \n$ and $\gamma = (1-v^2/c^2)^{-1/2}$. 
The \emph{usual} form of the relativistic Lagrangian is
\begin{equation}
L =  - m c^2 \sqrt{1-||\dot\x||^2/c^2},
\end{equation}
so
\begin{equation}
\p = {m \dot\x \over\sqrt{1-||\dot\x||^2/c^2}}; \qquad H = \p\cdot\dot\x - L = {mc^2\over \sqrt{1-||\dot\x||^2/c^2}}.
\end{equation}
Furthermore
\begin{equation}
\bar L =  - m c^2 \sqrt{1-||\dot{\bar\x}||^2/c^2},
\end{equation}
so
\begin{equation}
\bar\p = {m \dot{\bar\x} \over\sqrt{1-||\dot{\bar\x}||^2/c^2}}; \qquad 
\bar H = \bar\p\cdot\dot{\bar\x} - \bar L = {mc^2\over \sqrt{1-||\dot{\bar\x}||^2/c^2}},
\end{equation}
and in fact
\begin{equation}
\bar L \;d\bar t = L \; dt, 
\end{equation}
implying both $\epsilon=0$ and $\bpi=\0$, and $\bar\epsilon=0$ and $\bar\bpi=\0$. Then the energy-momentum transformations are just the usual \emph{linear} Lorentz transformations
\begin{equation}
(E,-\p^T) =  (\bar E, -\bar \p^T) M,
\end{equation}
and
\begin{equation}
(\bar E, -\bar \p^T) = (E,-\p^T) M^{-1}.
\end{equation}
This is the standard way of implementing Lagrangian and Hamiltonian mechanics in the presence of Lorentz symmetry. 

\subsection{Lorentz group (affine version)}

We could have chosen a slightly different normalization for $L$ and $H$. If we take
\begin{equation}
L =  m c^2  \left\{ 1-\sqrt{1-||\dot\x||^2/c^2}\right\},
\end{equation}
then
\begin{equation}
\p = {m \dot\x \over\sqrt{1-||\dot\x||^2/c^2}}; \qquad H = \p\cdot\dot\x - L = {mc^2\over \sqrt{1-||\dot\x||^2/c^2}} - mc^2.
\end{equation}
Furthermore
\begin{equation}
\bar L =  m c^2 \left\{ 1-\sqrt{1-||\dot{\bar\x}||^2/c^2}\right\},
\end{equation}
so
\begin{equation}
\bar\p = {m \dot{\bar\x} \over\sqrt{1-||\dot{\bar\x}||^2/c^2}}; \qquad 
\bar H = \bar\p\cdot\dot{\bar\x} - \bar L = {mc^2\over \sqrt{1-||\dot{\bar\x}||^2/c^2}} - mc^2.
\end{equation}
In fact with this normalization
\begin{equation}
[\bar L -mc^2] \;d\bar t = [L-mc^2] \;dt,
\end{equation}
whence
\begin{equation}
\bar\epsilon =  mc^2; \quad \bar\bpi=\0; \qquad\hbox{and}\qquad \epsilon= mc^2; \qquad \bpi = \0.
\end{equation}
We can rephrase this in terms of the 4-velocities of the ``aether'' and moving frames as
\begin{equation}
\left({\bar\epsilon\atop\bar\bpi}\right) = mc^2 \; \bar V; \qquad \left({\epsilon\atop\bpi}\right) = mc^2\; V;
\end{equation}
With these choices the energy-momentum transformations look slightly unusual. Taking $\v\parallel \p$ for simplicity (the non-collinear case does not teach us anything new) we now have
\begin{equation}
E \to \bar E  
= \gamma( [mc^2+ E]-\p\cdot \v)- mc^2,
\end{equation}
and
\begin{equation}
\p \to \bar \p 
= \gamma(\p-[mc^2+E]\v/c^2).
\end{equation}
The inverse transformations are 
\begin{equation}
\bar E \to E  
= \gamma([mc^2+\bar E] + \bar \p \cdot \v) - mc^2,
\end{equation}
and
\begin{equation}
\bar\p \to \p 
= \gamma ( [mc^2+\bar E]\v/c^2 +\bar \p).
\end{equation}
These \emph{affine} transformations make perfectly good physical sense once you realize that, with the conventions of this subsection, the $E$'s in question are just the relativistic kinetic energies --- quantities that are normally denoted by  $K$:
\begin{equation}
E_\mathrm{here} = E_\mathrm{total} - mc^2 = K.
\end{equation}
Then 
\begin{equation}
K \to \bar K  = \gamma( [mc^2+ K]-\p\cdot \v)- mc^2,
\end{equation}
and
\begin{equation}
\p \to \bar \p  = \gamma(\p-[mc^2+K]\v/c^2),
\end{equation}
while
\begin{equation}
\bar K \to K =    \gamma([mc^2+\bar K] + \bar \p \cdot \v) - mc^2,
\end{equation}
and 
\begin{equation}
\bar\p \to \p = \gamma ( [mc^2+\bar K]\v/c^2 +\bar \p).
\end{equation}
These are manifestly just a disguised form of the usual Lorentz transformations. 
Note that the formal $c\to \infty$ limit of \emph{these} (slightly nonstandard) affine equations is
\begin{equation}
K \to \bar K  =  K-\p\cdot \v +{1\over2}  m||\v||^2; \qquad
\p \to \bar \p  = \p-m\v;
\end{equation}
and
\begin{equation}
\bar K \to K =   \bar K + \bar \p \cdot \v  +{1\over2}  m||\v||^2; \qquad
\bar\p \to \p =\bar \p + m\v.
\end{equation}
These are the (affine) transformation laws for (the usual form of) the Galileo group. 

Again, the somewhat subtle message to take from this is that since the affine parameters $\epsilon$ and $\bpi$, and $\bar\epsilon$ and $\bar\bpi$, are already so important in situations of extremely high symmetry (the Lorentz group, the Galileo group), then they are also likely to be important in any situations where these symmetries are broken. 

\subsection{Galileo group (linear version)}

The previous discussion suggests that there might be \emph{some} (perhaps nonstandard) set of conventions that would make the energy and momentum transform linearly for the Galileo group. That is, there might be \emph{some} way of arranging things so that for the Galileo group
\begin{equation}
E \to \bar E = E - \p\cdot\v;
\qquad 
\p \to\bar \p = \p.
\end{equation}
How would we do that? It will have to be something rather unusual. 
Choose the following Lagrangians:
\begin{equation}
L = {1\over2} m ||\dot\x||^2;\qquad \bar L = {1\over2} m ||\dot{\bar\x} + \v||^2.
\end{equation}
Then the momenta are
\begin{equation}
\p = m \dot\x; \qquad \bar \p = m(\dot{\bar\x} + \v) =  m \dot\x = \p. 
\end{equation}
The energy in the aether frame is (as usual)
\begin{equation}
E = \p\cdot\dot \x - L = {1\over2}m ||\dot\x||^2.
\end{equation}
However with these conventions the energy in the moving frame is 
\begin{eqnarray}
\bar E = \bar\p\cdot\dot{\bar\x} - \bar L &=&  m(\dot{\bar\x}+\v)\cdot \dot{\bar\x} - {1\over2}m ||\dot{\bar\x}+\v||^2\\
&=& 
m\dot{\x}\cdot (\dot{\x}-\v) - {1\over2}m ||\dot{\x}||^2 
\\
&=&  {1\over2}m ||\dot{\x}||^2  - m \dot\x\cdot \v = E - \p \cdot \v. 
\end{eqnarray}
Now $\bar L = {1\over2} m ||\dot{\bar\x} + \v||^2$, is certainly an ``odd'' and ``unusual'' Lagrangian to choose for a free non-relativistic particle in the moving frame, but it is by no means ``wrong'' --- it certainly does the job. One certainly has the correct equations of motion $\ddot{\bar\x}=0$, and for \emph{this} definition  of energy and momentum, albeit ``odd'' and ``unusual'', the energy-momentum transformation laws are explicitly \emph{linear}:
\begin{equation}
\bar \p = \p; \qquad \bar E =  E - \p \cdot \v.
\end{equation}
Note that we have made the quantities $\{\epsilon,\bpi\}$ and $\{\bar\epsilon,\bar\bpi\}$  simple, in fact zero, at the price of making the moving-frame Lagrangian complicated. (For some comments in a similar vein, see section II.A of reference~\cite{arXiv:1110.1314}.)

\subsection{Summary}

When looking at how this general framework and formalism applies to the Lorentz group we saw that there were good choices for $\epsilon$ and $\bpi$, and  $\bar\epsilon$ and $\bar\bpi$, and also ``bad'' (or rather, sub-optimal) choices.  
There seems to be considerable freedom in how one picks $\epsilon$ and $\bpi$, and $\bar\epsilon$ and $\bar\bpi$, and so considerable freedom in choosing \emph{affine} versus \emph{linear} transformations for the 4-momentum. 
Can this freedom be used to improve things?
If one is working in a region of parameter space that is ``close'' to special relativity (a ``perturbative'' deviation from special relativity) then linear transformations for the 4-momentum would seem to be the most appropriate choice. 
If one is working in a region of parameter space that is ``close'' to Galillean  relativity (a ``perturbative'' deviation from Newtonian mechanics) then affine transformations for the 4-momentum would seem to be the most appropriate choice. 
The general situation is somewhat unclear, but it seems advisable to retain the generality of the full affine transformations as long as possible.

\section{On-shell energy-momentum relations}

In any particular inertial frame if one measures the energy $E$ and momentum $\p$ of an on-shell particle then there will be \emph{some} relation between them; an on-shell energy-momentum relation $E= E(\p)$. One normally expects a very tight connection between the functional form of these 
energy-momentum relations and the functional form of the transformations between inertial frames --- unfortunately this very tight connection is intimately related with adopting the relativity principle, and will in general fail once the relativity principle is abandoned. That is, in Lorentz violating theories the  functional form of the energy-momentum relations and the functional form of the transformations between inertial frames can be (and often are) independent of each other.

\subsection{Rest energy without the relativity principle}

To see how this comes about, consider the preferred (aether) frame $F$, and in that frame suppose you measure the energy $E$ and momentum $\p$ of the same particle in a number of different kinematic states to map out the energy-momentum relation $E= E(\p)$ in the aether frame.  To each momentum $\p$ we associate a 3-velocity $\v = \partial E/\partial\p$. Now go to the rest frame $\bar F$ of the particle (of course the rest frame of the particle is moving with respect to the aether). In the rest frame the particle will by definition have 3-velocity zero $\bar\v=0$, and will have some energy, call it the rest-energy $\bar E = E_0$ and some momentum, call it the rest-momentum $\bar\p = \p_0$. 

If the relativity principle holds then the rest-energy and rest-momentum must be intrinsic properties of the particle that cannot depend on its velocity with respect to the aether --- and in particular the rest-momentum is most typically chosen to be zero. But once one has preferred frame effects the rest-energy and rest-momentum can very definitely depend on the state of motion with respect to the aether. That is, generally we will have  $\bar E = E_0(\bar F)$ and $\bar\p = \p_0(\bar F)$.

Transforming back to the aether frame we now see
\begin{equation}
E =   \gamma E_0(\bar F) + \p_0(\bar F)^T \, \Sigma \v  + \gamma\bar \epsilon + \bar \bpi^T \, \Sigma \v  -\epsilon,
\end{equation}
and
\begin{equation}
\p = \gamma E_0(\bar F) \u + \Sigma^T \,\p_0(\bar F) + \gamma\bar \epsilon \u + \Sigma^T \bar \bpi  -\bpi.
\end{equation}
In general, unless further assumptions are made, this is the best one can do.

We now use the freedom to choose the quantities $\{\epsilon,\bpi\}$ and $\{\bar\epsilon,\bar\bpi\}$ to make life as simple as possible. 
Without any real loss of generality we can choose $\bar\bpi = - \p_0(\bar F)$ in which case
\begin{equation}
E =   \gamma E_0(\bar F) + \gamma\bar \epsilon -\epsilon,
\end{equation}
and
\begin{equation}
\p = \gamma E_0(\bar F) \u  + \gamma\bar \epsilon \u  -\bpi.
\end{equation}
(This is equivalent to choosing conventions so that in the rest frame the total ``effective'' rest momentum $\p_0+\bar\bpi=\0$.)
Let us now for definiteness choose $\epsilon= 0$ and $\bpi=\0$, then
\begin{equation}
E =   \gamma [E_0(\bar F)+\bar\epsilon],
\end{equation}
and
\begin{equation}
\p = \gamma [E_0(\bar F)+\bar\epsilon] \u =  E \u.
\end{equation}
(We have done things in this manner so that it becomes clear just how general the relation $\p = E\u$ really is.)
Finally choose $\bar\epsilon=0$, then with these choices we can write
\begin{equation}
E =   \gamma E_0(\bar F),
\end{equation}
and
\begin{equation}
\p = \gamma E_0(\bar F) \u = E \u.
\end{equation}
Introduce an arbitrary but fixed constant $c$ with the dimensions of velocity (not necessarily the speed of light), and some arbitrary function $\varpi(\bar F)$ which is completely at our disposal. Then in the aether frame we can write
\begin{equation}
E^2 - \varpi ||\p||^2 c^2 =  \gamma^2 (1- \varpi c^2 ||\u||^2) E^2_0(\bar F).
\end{equation}
Two particularly useful (but by no means inevitable) choices are to take:
\begin{itemize}
\item 
Choose $\varpi\to 1$ so that
\begin{equation}
E^2 - ||\p||^2 c^2 =  \gamma^2 (1- c^2 ||\u||^2) E^2_0(\bar F).
\end{equation}
\item
Choose $\varpi\to ||\v||^2/(||\u||^2 c^4)$ so that
\begin{equation}
E^2 -  {||\v||^2\over||\u||^2 c^2} \; ||\p||^2 =  \gamma^2 (1- v^2/c^2) E^2_0(\bar F).
\end{equation}
\end{itemize}

\noindent
In the case of exact Lorentz invariance we most usefully choose $\varpi\to1$, with the constant $c$ being interpreted as the speed of light, and $\gamma\to1/\sqrt{1-v^2/c^2}$, while $\u = \v/c^2$. Furthermore $E_0$ is then independent of $\v$, so in this case one recovers the usual kinematic relation $E^2 - ||\p||^2 c^2  = E_0^2$, while (as expected) $E=\gamma E_0$ and $\p = \gamma E_0\v/c^2 = E \v/c^2$.  In the absence of Lorentz invariance one generically has to live with the more complicated kinematics presented above. The notion of rest energy $E_0$ still makes perfectly good sense, but the rest energy can depend on the particle's state of motion through the aether, $E_0(\bar F)$, and the relation to 4-momentum is considerably more subtle than one might have expected.

The key point here is that the energy-momentum relation $E(\p)$ and the transformation matrix $M$ are in general independent of each other; knowing one does not necessarily give you the other (except when Lorentz invariance is assumed, or some similar restriction is imposed).  There are two additional special cases of considerable interest, which we now discuss.

\subsection{Invariant rest energy without the relativity principle}

One can speculate or hypothesize that for unknown reasons the internal structure of elementary and composite particles self-regulates so that rest energies are independent of one's state of motion through the aether. One still has rather unusual behaviour in that
\begin{equation}
E =   \gamma E_0; 
\qquad
\p = \gamma E_0 \u = E \u;
\end{equation}
while
\begin{equation}
E^2 - \varpi ||\p||^2 c^2 =  \gamma^2 (1- \varpi c^2 ||\u||^2) E^2_0.
\end{equation}
(Remember $\u$ is not necessarily parallel to $\v$, neither does $||\u||$ equal $||\v||$, they do not even have the same dimensions. In addition, all three of the functions $\varpi(\bar F)$, $\gamma(\bar F)$, and $\u(\bar F)$ can depend on the particle's state of motion with respect to the aether. In fact $\varpi$ is entirely arbitrary and can be adjusted to taste --- we will have cause to use this freedom below.)
So even with an invariant rest mass (and this is a rather strong assumption) the 4-momentum of a moving particle is rather definitely non-trivial.

\subsection{First minimalist Lorentz-violating model}

Another important special case to consider is to assume that the transformations between inertial frames are the usual Lorentz transformations but the energy-momentum relation for at least some of the particles is not Lorentz invariant. This is less bizarre than one might at first glance suspect, and is in fact the option that is in many ways most relevant to the OPERA--MINOS observations. The point is that the physical clocks and rulers we use in our laboratories have internal structures that are for all practical purposes independent of neutrino physics --- and we have good phenomenological/observational evidence that (apart possibly from the neutrino sector) Lorentz invariance is an extremely good approximation to empirical reality.  So it makes good sense to work in an approximation where all physical clocks and rulers are exactly Lorentz invariant, and the only Lorentz violations are hiding in the neutrino sector. In contrast, if there are significant Lorentz violations in the physics underpinning one's clocks and rulers, then using the Lorentz transformations to inter-relate the space and time coordinates determined by those clocks and rulers would be a very \emph{bad} and physically unjustified approximation.

More generally, stepping beyond the OPERA--MINOS scenario, let us consider a model where Lorentz-violating physics is confined to a specific sub-sector of the particle physics spectrum.
In this situation the rest energy of the Lorentz-violating particles can still depend on their state of motion with respect to the aether. 
In the aether frame we then have
\begin{equation}
E =   \gamma \,E_0(\bar F); \qquad  \p = \gamma \,E_0(\bar F) \, \v/c^2 = E\v/c^2;
\end{equation}
while
\begin{equation}
E^2 - ||\p||^2 c^2 =  E^2_0(\bar F).
\end{equation}
Again, even in this simplified situation, the 4-momentum and the kinematic relation are rather definitely non-trivial. If we now transform to a third inertial frame $\bar{\bar F}$, then certainly
\begin{equation}
\bar{\bar E}^2 - ||\bar{\bar\p}||^2 c^2 = E^2 - ||\p||^2 c^2 =  E^2_0(\bar F)
\end{equation}
is a Lorentz invariant, but the specific \emph{value} of this Lorentz invariant quantity depends on the absolute state of motion of the Lorentz-violating particle as viewed from the aether frame.  The way we have currently set things up, this rest energy could even be direction dependent --- no isotropy assumption (at least in the Lorentz-violating sector) has yet been made.
If we now add the additional assumption that the Lorentz violating physics is isotropic in the aether frame then
\begin{equation}
E =   \gamma E_0(v); \qquad  \p = \gamma E_0(v) \, \v/c^2 = E\v/c^2;
\end{equation}
while
\begin{equation}
E^2 - ||\p||^2 c^2 =  E^2_0(v).
\end{equation}
So we rather explicitly see the manner in which absolute speed with respect to the aether would formally affect on-shell particle energy-momentum relations.
We emphasise that in this model, even though the Lorentz-violating particles do not have a Lorentz invariant energy-momentum relation, their energies and 3-momenta nevertheless transform in the usual manner under Lorentz transformations. To make this look more relativistic, one  could  introduce a 4-velocity $V_\mathrm{aether}$ for the aether, and another 4-velocity $V_\mathrm{particle}$ for the Lorentz-violating particle. The speed $v$ of the  Lorentz violating particle with respect to the aether is then the usual  explicit function of the 4-inner-product $ \eta(V_\mathrm{aether},V_\mathrm{particle})$ and the kinematic relation takes the form 
\begin{equation}
E^2 - ||\p||^2 c^2 =  E^2_0( \eta(V_\mathrm{aether},V_\mathrm{particle})).
\end{equation}
This model is the first of the ``minimalist'' models of Lorentz violation we refer to in the abstract and introduction.  It is particularly useful in that, if one chooses to confine Lorentz violating physics to the neutrino sector, it gives one a very clean specific ``target'' to begin thinking about when analyzing the OPERA--MINOS observations. 

More broadly speaking, even beyond the OPERA--MINOS context, something along the lines of this minimalist model is often implicitly adopted in currently extant analyses of Lorentz violating models, but often without the relevant assumptions being clearly and explicitly laid out. Many current analyses implicitly treat Lorentz violation perturbatively, modelling reality by a Lorentz invariant ``core'' subject to Lorentz-violating ``perturbations''. (This is true for instance in the Kostelecky \emph{et al.}~Standard Model Extension~\cite{Colladay:1998fq, Kostelecky:1988zi, Kostelecky:2003fs, Kostelecky:2000mm, Kostelecky:2002hh, Kostelecky:2003cr, Kostelecky:1999mr, Kostelecky:2001mb, Bear:2000cd}, the Coleman--Glashow~analyses~\cite{Coleman:1997xq, Coleman:1998ti}, see also~\cite{Cohen:2011hx}, and the Jacobson--Liberati--Mattingly analyses~\cite{Liberati:2001cr, Jacobson:2001tu, Jacobson:2002hd, Mattingly:2002ba, Jacobson:2002ye, Jacobson:2003ty, Jacobson:2003bn, Jacobson:2004qt, Jacobson:2004rj, Jacobson:2005bg, Mattingly:2005re}, see also~\cite{Maccione:2011fr}.) Typically, to a first approximation one ignores the effect of any (presumably small) Lorentz-violating physics on the internal structure one's clocks and rulers, thereby implicitly justifying the use of ordinary Lorentz transformations for one's experimentally measured energy and momenta, and focuses attention on subtle deviations from Lorentz invariance that might be probed by suitably designed null experiments.  However it should be very much emphasised that if the effect of Lorentz-violating physics ever has non-perturbatively large effects on the internal structure one's clocks and rulers, then one can no longer safely use the ordinary Lorentz transformations for experimentally determined energy and momenta --- and instead of adopting some version of the minimalist model above one must then resort to the full power of the preceding analysis.

\subsection{Summary}

We emphasise that we have gone to all this trouble in setting up a very general formalism in order to have a coherent and consistent framework to operate in once we begin to entertain possible departures from Lorentz invariance.  Many of the results derived so far are quite unexpected when one has been trained to always think in a Lorentz invariant (or even Galilean invariant) and relativity principle respecting manner.

\section{Adding more constraints}

\subsection{Linearity plus isotropy}
Now let us add the assumption of isotropy --- specifically that physics is isotropic in the preferred frame, the aether frame. 
In particular this means that in the inertial transformation matrices
\begin{equation}
M = \left[\begin{array}{c|c} \vphantom{\Big|}\gamma & -\gamma\u^T \\ \hline 
\vphantom{\Big|}-\Sigma \v &  \Sigma \end{array} \right];
\qquad
M^{-1} = \left[
\begin{array}{c|c} \vphantom{\Big|} 
\gamma^{-1} (1-\u  \cdot\v)^{-1} & \u^T (I - \v\otimes\u)^{-1} \Sigma^{-1}\\ \hline 
\vphantom{\Big|}
\gamma^{-1} (1-\u \cdot\v)^{-1}  \v &  (I - \v\otimes\u)^{-1} \Sigma^{-1}
\end{array} 
\right];
\end{equation}
all vectors and matrices should be constructible only using the vector $\v$ and its magnitude --- there are now assumed to be no preferred principal axes for the universe. We are also assuming that the frames $F$ and $\bar F$ are ``aligned'' (not rotated with respect to each other). 
Then isotropy amounts to
\begin{equation}
\u || \v;  \qquad \Sigma =  a I + b \v \otimes \v.
\end{equation}
In fact it is now useful to introduce an arbitrary but fixed unspecified constant $c$ with the dimensions of velocity, and a dimensionless parameter $\zeta$, to write
\begin{equation}
\u = \zeta \v /c^2.
\end{equation}
Similarly, let us introduce dimensionless variables $\chi$ and $\xi$ to write
\begin{equation}
\Sigma = \gamma \chi \n \otimes \n + \xi[I-\n\otimes \n].
\end{equation}
Recall $\v = v \n$. By appealing to isotropy, the four quantities $\gamma$, $\chi$, $\zeta$, and $\xi$ are arbitrary dimensionless functions of the dimensionless variable $v^2/c^2$. By combining linearity with isotropy in this manner we have obtained a variant of the Robertson--Mansouri--Sexl framework; see~\cite{RMS1, RMS2}, and section 3.2 of~\cite{Mattingly:2005re}. (The RMS formalism invokes several other technical assumptions not relevant to the current discussion, and is not quite identical to our own framework.) Note that the quantities $\gamma$, $\chi$, $\zeta$, and $\xi$ can still depend on the internal structure of one's clocks and rulers.

We now have
\begin{equation}
M = \left[\begin{array}{c|c} \vphantom{\Big|}\gamma & -\gamma\zeta\v^T/c^2 \\ \hline 
\vphantom{\Big|}-\gamma \chi \v &  \gamma \chi \n \otimes \n + \xi[I-\n\otimes \n] \end{array} \right].
\end{equation}
An intermediate step in calculating the inverse transformation is
\begin{equation}
M^{-1} = \left[
\begin{array}{c|c} \vphantom{\Big|} 
\gamma^{-1} (1-\zeta v^2/c^2)^{-1} & \zeta \v^T (I - \zeta \v \v^T/c^2)^{-1} \Sigma^{-1}/c^2\\ \hline 
\vphantom{\Big|}
\gamma^{-1} (1-\zeta v^2/c^2)^{-1}  \v &  (I - \zeta \v \v^T/c^2)^{-1} \Sigma^{-1}
\end{array} 
\right].
\end{equation}
But
\begin{eqnarray}
\Sigma (I - \zeta \v \v^T/c^2) &=& (\gamma \chi \n \otimes \n + \xi[I-\n\otimes \n])(I - [\zeta v^2/c^2]\n\otimes\n) 
\nonumber
\\
&=& \gamma \chi [1-\zeta v^2/c^2] \n \otimes \n + \xi[I-\n\otimes \n],
\end{eqnarray}
whence
\begin{equation}
(I - \zeta \v \v^T/c^2)^{-1} \Sigma^{-1} = \gamma^{-1} \chi^{-1} [1-\zeta v^2/c^2]^{-1} \n \otimes \n + \xi^{-1}[I-\n\otimes \n].
\end{equation}
So the inverse transformation matrix simplifies to
\begin{equation}
M^{-1} = \left[
\begin{array}{c|c} \vphantom{\Big|} 
\gamma^{-1} (1-\zeta v^2/c^2)^{-1} & \gamma^{-1} (1 - \zeta v^2/c^2)^{-1}  \zeta \chi^{-1}  \v^T/c^2\\ \hline 
\vphantom{\Big|}
\gamma^{-1} (1-\zeta v^2/c^2)^{-1}  \v &  \gamma^{-1} \chi^{-1} [1-\zeta v^2/c^2]^{-1} \n \otimes \n + \xi^{-1}[I-\n\otimes \n]
\end{array} 
\right].
\end{equation}
By a specialization of our previous discussion:
\begin{itemize}
\item The velocity of the moving frame with respect to the aether is $\v$.
\item The velocity of the aether with respect to the moving frame is $-\chi \v$. 
\item 
These are now at least collinear, and in fact anti-parallel, but can still differ in magnitude; they are still not equal-but-opposite.
\end{itemize}
If we rotate to align $\v$ along the $\hat\x$ axis this looks a little simpler:
\begin{equation}
M = \left[\begin{array}{c|c|c} \vphantom{\Big|}\gamma & -\gamma\zeta v/c^2 &\0^T \\ \hline 
\vphantom{\Big|}-\gamma \chi v &  \gamma \chi &\0^T\\ \hline\
\0 & \0 & \xi I
\end{array} \right],
\end{equation}
and
\begin{equation}
M^{-1} = \left[
\begin{array}{c|c|c} \vphantom{\Big|} 
\gamma^{-1} (1-\zeta v^2/c^2)^{-1} & \gamma^{-1} (1 - \zeta v^2/c^2)^{-1}  \zeta \chi^{-1}  v/c^2 &\0^T\\ \hline 
\vphantom{\Big|}
\gamma^{-1} (1-\zeta v^2/c^2)^{-1}  v &  \gamma^{-1} \chi^{-1} [1-\zeta v^2/c^2]^{-1}  &\0^T \\
\hline
\0 & \0 & \xi^{-1} I
\end{array} 
\right].
\end{equation}
This is as far as you can get with linearity plus isotropy --- you still have four arbitrary functions $\gamma(v^2/c^2)$, $\chi(v^2/c^2)$, $\zeta(v^2/c^2)$, and $\xi(v^2/c^2)$ to deal with, but at least it is no longer an arbitrary $4\times 4$ matrix with 16 free components.  The set of transformations is still not a group, just a groupoid/pseudogroup.  

In view of the isotropy assumption particle rest masses $E_0$ should depend only on the speed with respect to the aether, hence be of the form $E_0(v)$. Specializing our earlier discussion, with $\varpi$, $\gamma$, and $\zeta$ being velocity dependent, in the aether frame we would have
\begin{equation}
E = \gamma E_0(v); \qquad \p = \gamma \zeta E_0(v) \v/c^2;
\end{equation}
with
\begin{equation}
E^2 - \varpi ||\p||^2 c^2 = \gamma^2 \left[1-\varpi\zeta^2 v^2/c^2\right] E_0(v)^2.
\end{equation}
Note that Lorentz invariance corresponds to setting $\chi=\zeta=\xi=1$,  with $\gamma=1/\sqrt{1-v^2/c^2}$, (and in addition demanding $\varpi\to 1$ and that $E_0$ be constant).

The Galilean limit is somewhat delicate: Physically we want to be looking at some sort of low velocity limit. When moving at zero velocity through the aether we expect $M\to I$ (corresponding to the trivial transformation),  so we must have $\chi(0)=\gamma(0)=\xi(0)=1$. In contrast $\zeta(0)$ should be finite but is otherwise unconstrained. However $c$ is at this stage just some constant with the dimensions of velocity, it does not yet have any deeper physical interpretation, so one can simply absorb $\zeta(0)$ into a redefinition of $c$ and so effectively set $\zeta(0)\to1$.
\begin{itemize}
\item In the transformation matrices $M$ and $M^{-1}$, this low-velocity limit corresponds to $\zeta=\chi=\gamma=\xi=1$, with $||\v|| \ll c$.
\item
Because of isotropy, in the low-velocity limit we must have both
\begin{equation}
\gamma(v)\approx 1 + {1\over2}\gamma_2 v^2/c^2 +\dots, 
\quad\hbox{and}\quad
\zeta(v)\approx 1 + {1\over2}\zeta_2 v^2/c^2 +\dots.
\end{equation}
Furthermore
\begin{equation}
 E_0(v)= E_0(0) \; \left\{1 + {1\over2}\kappa_2  v^2/c^2+\dots \right\},
\end{equation}
so that:
\begin{equation}
E \approx E_0(0) + {1\over2}[E_0(0)/c^2] \{\gamma_2 + \kappa_2\} v^2 + \dots;
\qquad
\p \approx [E_0(0)/c^2] \v + \dots
\end{equation}
\item
If we define the low-velocity effective mass by $m_\mathrm{eff}= E_0(0)/c^2$ then
\begin{equation}
E \approx m_\mathrm{eff} c^2 +  \{\gamma_2 + \kappa_2\} \; { ||\p||^2 \over 2 m_\mathrm{eff}} + \dots;
\qquad
\p \approx  m_\mathrm{eff}\; \v + \dots
\end{equation}

\end{itemize}
So there is a sensible low-velocity limit, though it is perhaps more subtle than one might have thought.

\subsection{Linearity plus isotropy plus reciprocity}

It is sometimes useful to restrict attention to situations where $M^{-1}(\v) = M(-\v)$. Note that this is an additional axiom beyond homogeneity and isotropy. 
\begin{itemize}
\item  This is (one version of) the so-called \emph{reciprocity principle}.  It is still weaker than the relativity principle. 
\item This version of the reciprocity principle, because it also makes assumptions about the transverse directions,  is very slightly stronger than asserting that the velocity of any inertial frame as seen from the aether is minus the velocity of the aether as seen from that inertial frame~\cite{Berzi:1969}.
\item The way we have formulated it, reciprocity implies both $\chi=1$ and $\xi=1$, and in addition imposes the constraint 
\begin{equation}
\gamma = {1\over\sqrt{1-\zeta v^2/c^2}}.
\end{equation}
\end{itemize}
To see this, compare $M$ with $M^{-1}$ above, and note that $M^{-1}(\v) = M(-\v)$ implies the three relations:
\begin{equation}
\gamma = \gamma^{-1} (1-\zeta v^2/c^2)^{-1};
\end{equation}
\begin{equation}
\gamma \chi = \gamma^{-1} \chi^{-1} (1-\zeta v^2/c^2)^{-1};
\end{equation}
\begin{equation}
\xi = \xi^{-1}.
\end{equation}
Solving, we see
\begin{equation}
\xi = 1; \qquad \chi = 1; \qquad  \gamma^2 (1-\zeta v^2/c^2) = 1.
\end{equation}
Then
\begin{equation}
M = \left[\begin{array}{c|c|c} \vphantom{\Big|}\gamma & -\gamma\zeta v/c^2 &\0^T \\ \hline 
\vphantom{\Big|}-\gamma v &  \gamma &\0^T\\ \hline\
\0 & \0 &  I
\end{array} \right];
\qquad
M^{-1} = \left[
\begin{array}{c|c|c} \vphantom{\Big|} 
\gamma & \gamma \zeta  v/c^2 &\0^T\\ \hline 
\vphantom{\Big|}
\gamma v &  \gamma  &\0^T \\
\hline
\0 & \0 & I
\end{array} 
\right];
\end{equation}
subject to the constraint
\begin{equation}
\gamma = {1\over\sqrt{1-\zeta v^2/c^2}}.
\end{equation}
Note that you now only have one free function $\zeta(v^2/c^2)$, everything else is determined.
\begin{itemize}
\item 
Working along a somewhat different route, it has been shown~\cite{Berzi:1969} that combining relativity+homogeneity+isotropy implies (at least one version of) the reciprocity principle. 

\item
Note that adopting the principle of reciprocity implies the set $\{M(\v)\}$ is now closed under matrix inversion, though it is still not a group.

\item 
This is not quite special relativity [or even Galilean relativity], but it is getting awfully close.
\end{itemize}
%
\subsection{Second minimalist Lorentz-violating model}

Since the model above (linearity plus isotropy plus reciprocity)  is a simple one-function violation of special relativity, it holds a special place in the set of all Lorentz violating (relativity principle violating) theories --- this is arguably the simplest  violation of special relativity one can have \emph{at the level of the transformations between inertial frames}.  At the level of the coordinate transformations
\begin{equation}
t \to \bar t = {t-\zeta(v) v x/c^2\over\sqrt{1-\zeta(v) v^2/c^2}};
\end{equation}
\begin{equation}
x \to \bar x = {x - vt \over\sqrt{1-\zeta(v) v^2/c^2}};
\end{equation}
\begin{equation}
y \to \bar y = y; \qquad z \to \bar z = z.
\end{equation}
The closest one can get to a notion of ``interval'' is to observe
\begin{equation}
{c^2 (\Delta t)^2\over \zeta(v)} - ||\Delta\x||^2 = {c^2 (\Delta\bar t\,)^2\over\zeta(v)} - ||\Delta\bar\x||^2.
\end{equation}
Recall that $\zeta(v)$ depends on the absolute speed of the moving frame through the aether, so this is only a 2-frame invariant, it is not a general invariant for arbitrary combinations of inertial frames. To be explicit about this, let $F_1$ and $F_2$ be two moving frames, then
\begin{equation}
{c^2 (\Delta t)^2\over \zeta(v_1)}- ||\Delta\x||^2 = {c^2 (\Delta t_1\,)^2\over \zeta(v_1)} - ||\Delta\x_1||^2,
\end{equation}
and
\begin{equation}
{c^2 (\Delta t)^2\over \zeta(v_2)}- ||\Delta\x||^2 = {c^2 (\Delta t_2\,)^2\over \zeta(v_2)} - ||\Delta\x_2||^2.
\end{equation}
But (ultimately due to the lack of the relativity principle, and the consequent lack of group structure for the transformations) there is, under the current assumptions, no simple relationship of this type connecting the measurements in inertial frame $F_1$ with those in inertial frame $F_2$. 

When we turn to on-shell particle energy-momentum relations we still have invariant masses $E_0(v)$ that can depend on absolute velocity with respect to the aether. Therefore, in view of our previous discussion,   in the aether frame we would have
\begin{equation}
E = {E_0(v)\over \sqrt{1-\zeta(v) v^2/c^2}} ; \qquad \p = {\zeta(v) E_0(v) \v/c^2\over\sqrt{1-\zeta(v) v^2/c^2}}  = E \v /c^2;
\end{equation}
with
\begin{equation}
E^2 - \varpi(v) ||\p||^2 c^2 =  \left[{1-\varpi(v) \zeta^2(v) v^2/c^2\over 1 - \zeta(v) v^2/c^2}\right] E_0(v)^2.
\end{equation}
But $\varpi(v)$ is a completely arbitrary function that is entirely at our disposal, so in the current context it makes sense to choose $\varpi=1/\zeta$ in which case
\begin{equation}
E^2 - \zeta^{-1}(v) \, ||\p||^2 c^2 =  E_0(v)^2.
\end{equation}
Even if we make the additional and rather stringent assumption that rest masses are invariant, independent of absolute velocity through the aether, (and this is very definitely an extra assumption beyond reciprocity), one still picks up non-trivial physics via the $v$-dependent function $\zeta(v)$:
\begin{equation}
E = {E_0\over \sqrt{1-\zeta(v) v^2/c^2}} ; \qquad \p = {\zeta(v) E_0 \v/c^2\over\sqrt{1-\zeta(v) v^2/c^2}}  = E \v /c^2;
\end{equation}
with
\begin{equation}
E^2 -\zeta^{-1}(v)\, ||\p||^2 c^2 = E_0^2.
\end{equation}
This model is the second of the ``minimalist'' models of Lorentz violation we refer to in the abstract and introduction.  It is particularly useful in that it gives one a very clean specific ``target'' to take aim at. 

\subsection[Linearity plus isotropy plus reciprocity plus relativity]{Linearity plus isotropy plus reciprocity plus relativity}

If we now (finally) adopt the relativity principle, then for arbitrary $\v_1$ and $\v_2$ the object $M(\v_2,\v_1)$ must equal $M(\w)$ for \emph{some} $\w(\v_1,\v_2)$ (with $\w$ being interpreted as the relative velocity of the two inertial frames).  But this then implies that the set $\{ M(\v)\}$ forms a group, not merely a groupoid/pseudogroup. We shall see that this group condition implies $\zeta=1$, whence finally $\gamma = 1/\sqrt{1-v^2/c^2}$. But $c$ was some arbitrary quantity with the dimensions of velocity, it was not pre-judged to be the physical speed of light.  Finite $c$ gives you the Lorentz group, infinite $c$ gives the Galileo group. (And the exceptional case $c^2<0$ actually means one is in Euclidean signature, and one obtains the $SO(4)$ rotation group. This exceptional case is normally excluded by appeal to a ``pre-causality'' principle~\cite{gr-qc/0107091}.)

As an explicit check, assuming linearity+isotropy+reciprocity we have
\begin{equation}
M_1 = \left[\begin{array}{c|c|c} \vphantom{\Big|}\gamma_1 & -\gamma_1\zeta_1 v_1/c^2 &\0^T\\ \hline 
\vphantom{\Big|}-\gamma_1 v_1 &  \gamma_1 &\0^T\\ \hline\
\0 & \0 &  I
\end{array} \right];
\qquad
M_2 = \left[\begin{array}{c|c|c} \vphantom{\Big|}\gamma_2 & -\gamma_2\zeta_2 v_2/c^2 &\0^T \\ \hline 
\vphantom{\Big|}-\gamma_2 v_2 &  \gamma_2 &\0^T\\ \hline\
\0 & \0 &  I
\end{array} \right];
\end{equation}
subject to the constraint
\begin{equation}
\gamma_i = {1\over\sqrt{1-\zeta_i v_i^2/c^2}}.
\end{equation}
Then the group property requires the existence of some $v_{12}$ such that
\begin{equation}
M_1 M_2  = M_{12}.
\end{equation}
Explicitly:
\begin{equation}
 \left[\begin{array}{c|c|c} \vphantom{\Big|}\gamma_1\gamma_2(1+\zeta_1v_1v_2/c^2) &
 -\gamma_1\gamma_2 (\zeta_1 v_1+\zeta_2 v_2) /c^2 &\0^T \\ \hline 
\vphantom{\Big|}-\gamma_1\gamma_2( v_1 +v_2) &
  \gamma_1\gamma_2(1+\zeta_2v_1v_2/c^2)  &\0^T\\ \hline\
\0 & \0 &  I
\end{array} \right]
=
 \left[\begin{array}{c|c|c} \vphantom{\Big|}\gamma_{12} & -\gamma_{12}\zeta_{12} v_{12} /c^2 &\0^T \\ \hline 
\vphantom{\Big|}-\gamma_{12} v_{12} &  \gamma_{12} &\0^T\\ \hline\
\0 & \0 &  I
\end{array} \right].
\end{equation}
But by comparing the diagonal elements this can be true only if $\zeta_1=\zeta_2$ for all values of $v_1$ and $v_2$. That is,  $\zeta(v_1)=\zeta(v_2)$ for all values of $v_1$ and $v_2$, so that 
there exists some \emph{velocity independent constant} $\zeta_0$ such that
\begin{equation}
\zeta(v)= \zeta_0.
\end{equation}
This now implies
\begin{equation}
M_1 = \left[\begin{array}{c|c|c} \vphantom{\Big|}\gamma_1 & -\gamma_1\zeta_0 v_1/c^2 &\0^T \\ \hline 
\vphantom{\Big|}-\gamma_1 v_1 &  \gamma_1 &\0^T\\ \hline\
\0 & \0 &  I
\end{array} \right];
\qquad
M_2 = \left[\begin{array}{c|c|c} \vphantom{\Big|}\gamma_2 & -\gamma_2\zeta_0 v_2/c^2 &\0^T \\ \hline 
\vphantom{\Big|}-\gamma_2 v_2 &  \gamma_2 &\0^T\\ \hline\
\0 & \0 &  I
\end{array} \right].
\end{equation}
The statement $M_1 M_2 = M_{12}$ becomes
\begin{equation}
\left[\begin{array}{c|c|c} \vphantom{\Big|}\gamma_1\gamma_2(1+\zeta_0v_1v_2/c^2) &
 -\gamma_1\gamma_2 \zeta_0 (v_1+v_2) /c^2 &\0^T \\ \hline 
\vphantom{\Big|}-\gamma_1\gamma_2( v_1 +v_2) &
  \gamma_1\gamma_2(1+\zeta_0v_1v_2/c^2)  &\0^T\\ \hline\
\0& \0 &  I
\end{array} \right]
=
 \left[\begin{array}{c|c|c} \vphantom{\Big|}\gamma_{12} & -\gamma_{12}\zeta_{0} v_{12} /c^2 &\0^T \\ \hline 
\vphantom{\Big|}-\gamma_{12} v_{12} &  \gamma_{12} &\0^T\\ \hline\
\0 & \0 &  I
\end{array} \right].
\end{equation}
But now we can simply absorb $\zeta_0$ into a redefinition of $c$. After all, $c$ is at this stage just an arbitrary but fixed constant with the dimensions of velocity. Taking $c^2\to c^2/\zeta_0$ we have
\begin{equation}
M_1 = \left[\begin{array}{c|c|c} \vphantom{\Big|}\gamma_1 & -\gamma_1v_1/c^2 &\0^T \\ \hline 
\vphantom{\Big|}-\gamma_1 v_1 &  \gamma_1 &\0^T\\ \hline\
\0 & \0 &  I
\end{array} \right];
\qquad
M_2 = \left[\begin{array}{c|c|c} \vphantom{\Big|}\gamma_2 & -\gamma_2 v_2/c^2 &\0^T \\ \hline 
\vphantom{\Big|}-\gamma_2 v_2 &  \gamma_2 &\0^T\\ \hline\
\0 & \0 &  I
\end{array} \right];
\end{equation}
\begin{equation}
M_1 M_2 = \left[\begin{array}{c|c|c} \vphantom{\Big|}\gamma_1\gamma_2(1+v_1v_2/c^2) &
 -\gamma_1\gamma_2 (v_1+v_2) /c^2 &\0^T \\ \hline 
\vphantom{\Big|}-\gamma_1\gamma_2( v_1 +v_2) &
  \gamma_1\gamma_2(1+v_1v_2/c^2)  &\0^T \\ \hline\
\0 & \0 &  I
\end{array} \right]
=
 \left[\begin{array}{c|c|c} \vphantom{\Big|}\gamma_{12} & -\gamma_{12} v_{12} /c^2 &\0^T \\ \hline 
\vphantom{\Big|}-\gamma_{12} v_{12} &  \gamma_{12} &\0^T\\ \hline\
\0 & \0 &  I
\end{array} \right].
\end{equation}
If $c^2$ is finite and positive, we have the Lorentz transformations. If $c^2$ is infinite we have Galileo's transformations. This is (essentially) von Ignatowsky's result. (Note that $c^2=0$ is hopelessly diseased,\footnote{It is at this stage, setting $c^2\to0$, that one could if desired obtain Carroll kinematics~\cite{Carroll1, dyson, hep-th/0410086, Carroll2, Carroll3} by enforcing the particular limit $c^2\to0$, while $v\to 0$, but holding the slowness $u=v/c^2$ fixed. The relevance to ``real world'' physics seems somewhat tenuous.}  while $c^2<0$ actually corresponds to Euclidean signature spacetime, with the set $\{M\}$ being the group $SO(4)$ of Euclidean rotations.) 

\section{Conclusions}

We have seen that once one for any reason moves away from Lorentz invariance, and specifically once one discards the relativity principle, then many of the intuitions one has been trained to develop in a special relativistic setting need to be significantly and carefully revised. In a companion article we had considered threshold phenomena~\cite{arXiv:1111.6340}, which can be studied by picking and working in a particular arbitrary but fixed inertial frame. In the current article we have carefully analyzed what happens to the transformation properties between inertial frames once the relativity principle is abandoned. A key message to take from the above is that the situation is not hopeless --- even in the absence of a relativity principle quite a lot can still be said regarding the transformation properties between inertial frames, the combination of 3-velocities, the transformation of 4-momenta, and the interplay between the energy-momentum relations for on-shell particles and the transformation properties between inertial frames.

Key features of the analysis are the groupoid/pseudo-group structure of the set of transformations, the fact that 4-momentum transforms affinely as a dual vector, the fact that there are a number of distinct stages by which Lorentz invariance can be recovered --- by successively imposing linearity, then isotropy, then reciprocity, and finally the relativity principle. The net result is a  coherent framework within which Lorentz symmetry breaking can be explored in a controlled and internally consistent manner, while retaining usual notions of local physics.  Overall this article, and the companion paper~\cite{arXiv:1111.6340}, provide general techniques of interest when analyzing the OPERA--MINOS (and related) observations; but are in no sense dependent on the details of those specific observations --- these articles provide general techniques of interest for handling large classes of Lorentz-violating but local physical models.

\clearpage
\section*{Acknowledgments} 

This research was supported by the Marsden Fund, 
administered by the Royal Society of New Zealand.  Additionally, Valentina Baccetti acknowledges support by a Victoria University PhD scholarship, and Kyle Tate acknowledges support by a Victoria University MSc scholarship.

\appendix
\section{Some matrix identities}
\label{A:identities}

Herein we collect some useful matrix identities of a purely technical nature.  First note that
\begin{equation}
(I - \v\otimes\u)^{-1} = I + \sum_{n=1}^\infty (\v\otimes \u)^n =  I + (\v\otimes \u) \sum_{n=1}^\infty (\u\cdot\v)^{n-1} 
=  I + {\v\otimes \u\over 1- \u\cdot \v},
\end{equation}
with this particular derivation holding for $|\u\cdot\v|<1$, though the result itself 
\begin{equation}
(I - \v\otimes\u)^{-1} 
=  I + {\v\otimes \u\over 1- \u\cdot \v},
\end{equation}
holds for $\u\cdot\v \neq1$, as can easily be verified by multiplying both sides of the equation above by $(I - \v\otimes\u)$ and noting that $\det(I - \v\otimes\u) = 1-\vv\cdot\u$. (The case  $\u\cdot\v =1$ is the kinematic singularity alluded to previously.) 
Therefore 
\begin{equation}
\u^T (I - \v\otimes\u)^{-1} = \u^T + {(\u\cdot \v) \u^T\over 1- \u\cdot \v} = { \u^T\over 1- \u\cdot \v},
\end{equation}
at least for $\u\cdot\v \neq1$. 
Similarly
\begin{equation}
(1- \u\cdot \v)(I - \v\otimes\u)^{-1} 
=  (1- \u\cdot \v) I + \v\otimes \u,
\end{equation}
for $\u\cdot\v \neq1$. 
Secondly observe
\begin{eqnarray}
(I-\v\otimes\u) (I-\dot\x\otimes\u)^{-1} 
&=& (I-\v\otimes\u) \left(I+ \sum_{n=1}^\infty (\dot\x\otimes\u)^n \right)
\nonumber
\\
&=& I-\v\otimes\u \left(\sum_{n=0}^\infty (\dot\x\cdot\u)^n \right) + \dot\x\otimes\u \left(\sum_{n=0}^\infty (\dot\x\cdot\u)^n \right) 
\nonumber
\\
&=& I - {\v\otimes\u \over 1- \dot\x\cdot\u} + {\dot\x\otimes\u  \over 1- \dot\x\cdot\u} 
\nonumber
\\
 &=& I + {(\dot \x - \v)\otimes \u\over 1- \dot\x\cdot\u},
\end{eqnarray}
with this particular derivation holding for $|\dot\x\cdot\u|<1$, though the result itself holds for $\dot \x\cdot\u \neq1$. 
Therefore, for $\dot \x\cdot\u \neq1$, we have
\begin{eqnarray}
(I-\v\otimes\u) (I-\dot\x\otimes\u)^{-1} (\dot\x-\v) &=& \left(  I + {(\dot \x - \v)\otimes \u\over 1- \dot\x\cdot\u} \right) (\dot\x-\v)
\nonumber
\\
&=& (\dot\x-\v) + {(\dot\x-\v) \u \cdot (\dot\x-\v)\over 1- \dot\x\cdot\u}
\nonumber
\\
&=& (\dot\x-\v) \left\{  1- \dot\x\cdot\u + \u \cdot (\dot\x-\v) \over 1- \dot\x\cdot\u \right\} 
\nonumber
\\
&=& (\dot\x-\v) \left\{  1-  \u \cdot \v \over 1- \dot\x\cdot\u \right\}.
\end{eqnarray}
 
\section{Consistency of dynamics and kinematics}
\label{A:consistency}

Note that from Hamilton's equations we know $\dot \x = \partial H/\partial\p$, 
so to first order (which is all we require) $\Delta E = \dot\x \cdot \Delta\p$. Then from our discussion of the energy-momentum transformation laws, and specifically the fact that energy-momentum differences transform linearly, we have
\begin{equation}
\dot\x \cdot \Delta\p =    \gamma\;\dot{\bar\x}\cdot\Delta\bar\p + \Delta \bar \p^T \Sigma \v =  
\Delta \bar\p \cdot (\gamma \dot{\bar\x} + \Sigma \v ),
\end{equation}
and
\begin{equation}
\Delta \p =   \gamma\; (\dot{\bar\x}\cdot\Delta\bar\p ) \u + \Sigma^T \Delta \bar \p  = 
 (\gamma \; \u\otimes \dot{\bar\x} +\Sigma^T) \Delta \bar\p. 
\end{equation}
But then, for arbitrary $\Delta\bar\p$
\begin{equation}
\left\{  \dot\x^T   (\gamma \; \u\otimes \dot{\bar\x} +\Sigma^T)-  (\gamma \dot{\bar\x}^T + \Sigma \v^T ) \right\}\Delta\bar\p  = 0,
\end{equation}
implying
\begin{equation}
 \dot\x^T   (\gamma \; \u\otimes \dot{\bar\x} +\Sigma^T) =  (\gamma \dot{\bar\x}^T + \v^T \Sigma^T ).
\end{equation}
That is
\begin{equation}
  (\gamma \;  \dot{\bar\x} \otimes \u +\Sigma) \dot\x   =  (\gamma \dot{\bar\x} + \Sigma \v )
\end{equation}
whence 
\begin{equation}
\dot\x =   (\gamma \; \dot{\bar\x} \otimes \u +\Sigma) ^{-1} (\gamma \dot{\bar\x} + \Sigma \v ).
\end{equation}
This is equivalent to the velocity transformation law we previously derived. (Note that $\dot{\bar \x}=\0$ implies $\dot\x = \v$, while $\dot \x=\0$ implies $\dot{\bar\x} = - \Sigma \v /\gamma$.)

It is perhaps easier to start from the inverse transformations
\begin{equation}
\Delta E \to \Delta \bar E =     {\Delta E- \Delta\p\cdot \v\over \gamma(1-\u\cdot \v)},
\end{equation}
and
\begin{equation}
\Delta\p \to \Delta\bar \p = (\Sigma^{-1})^T (I-\u\otimes\v)^{-1} (\Delta\p-\Delta E\u).
\end{equation}
The energy transformation equation implies
\begin{equation}
\dot{\bar\x}\cdot\Delta \bar\p = {(\dot\x-\v) \cdot \Delta\p \over \gamma(1-\u\cdot\v)},
\end{equation}
while the momentum transformation equation yields
\begin{eqnarray}
\Delta\bar \p &=&   (\Sigma^{-1})^T (I-\u\otimes\v)^{-1} (\Delta\p- [\dot \x \cdot \Delta\p]\u)
\\
&=&  (\Sigma^{-1})^T (I-\u\otimes\v)^{-1}  (I-\u\otimes \dot\x) \Delta\p. 
\end{eqnarray}
But then
\begin{equation}
\left\{ 
\dot{\bar\x}^T  (\Sigma^{-1})^T (I-\u\otimes\v)^{-1}  (I-\u\otimes \dot\x) -  {(\dot\x-\v)^T \over \gamma(1-\u\cdot\v)} 
\right\} \Delta\p = 0,
\end{equation}
whence
\begin{equation}
\dot{\bar\x}^T  (\Sigma^{-1})^T (I-\u\otimes\v)^{-1}  (I-\u\otimes \dot\x)  =  {(\dot\x-\v)^T \over \gamma(1-\u\cdot\v)}.
\end{equation}
Therefore
\begin{equation}
(I-\dot\x\otimes\u)(I-\v\otimes\u)^{-1} \Sigma^{-1}  \dot{\bar\x} =  {\dot\x-\v \over \gamma(1-\u\cdot\v)},
\end{equation}
and we see
\begin{equation}
\dot{\bar\x} =   \Sigma (I-\v\otimes\u) (I-\dot\x\otimes\u)^{-1}{(\dot\x-\v) \over \gamma(1-\u\cdot\v)}. 
\end{equation}
But (see appendix \ref{A:identities}) 
\begin{eqnarray}
(I-\v\otimes\u) (I-\dot\x\otimes\u)^{-1} 
 &=& I + {(\dot \x - \v)\otimes \u\over 1- \dot\x\cdot\u}.
\end{eqnarray}
Furthermore (see appendix \ref{A:identities}) 
\begin{eqnarray}
(I-\v\otimes\u) (I-\dot\x\otimes\u)^{-1} (\dot\x-\v) 
&=& (\dot\x-\v) \left\{  1-  \u \cdot \v \over 1- \dot\x\cdot\u \right\}.
\end{eqnarray}
So finally
\begin{equation}
\dot{\bar\x} =    {\Sigma(\dot\x-\v) \over \gamma(1-\u\cdot\dot\x)},
\end{equation}
which is the 3-velocity transformation law we had previously derived. (Note that $\dot{\bar \x}=\0$ implies $\dot\x = \v$, while $\dot \x=\0$ implies $\dot{\bar\x} = - \Sigma \v /\gamma$.)  This verifies the internal consistency of the manner in which our Hamiltonian/Lagrangian mechanics interacts with the generic transformation laws between inertial frames.



\end{document}